\documentclass[english]{article}

\usepackage{amsmath}
\usepackage{amsfonts}
\usepackage[bookmarks=true, bookmarksnumbered=true, allbordercolors={1 1 1}]{hyperref}
\usepackage{float}
\usepackage{placeins}
\usepackage[cal=boondoxo]{mathalfa}
\usepackage{tikz}
\usetikzlibrary{fit,positioning}
\usepackage{lscape}
\usepackage{longtable}
\usepackage{epstopdf}
\usepackage{float}
\usepackage{setspace}
\usepackage{textcomp}
\usepackage[title]{appendix}
\usepackage{booktabs}

\usepackage{algorithm}
\usepackage[end]{algpseudocode}
\algnewcommand\algorithmicto{\textbf{to }}
\usepackage{etoolbox}
\patchcmd{\thebibliography}{\section*{\refname}}{}{}{}

\usepackage{geometry}
\usepackage{natbib}

\let\oldabstract\abstract
\let\oldendabstract\endabstract
\makeatletter
\renewenvironment{abstract}
{%
               {\list{}{\addtolength{\leftmargin}{2em} % change this value to add or remove length to the the default
                        \listparindent 1.5em%
                        \itemindent    \listparindent%
                        \rightmargin   \leftmargin%
                        \parsep        \z@ \@plus\p@}%
                \item\relax}%
               {\endlist}%
\oldabstract}
{\oldendabstract}

\usepackage{bm}

\providecommand{\tabularnewline}{\\}

\makeatother

\usepackage{babel}
\begin{document}

\title{{\LARGE Inducing Sparsity and Shrinkage in Time-Varying Parameter Models}}
%\author[1]{Florian Huber\thanks{Corresponding author. Email: \href{mailto: florian.huber@sbg.ac.at}{florian.huber@sbg.ac.at}. Any views expressed in this paper represent those of the authors only and not necessarily  of the ECB or the Eurosystem.  We are grateful to Niko Hauzenberger and Michael Pfarrhofer for helpful comments and suggestions. The first author acknowledges funding from the Austrian Science Fund (FWF): ZK 35.}}
%\author[2]{Gary Koop}
%\author[3]{Luca Onorante}

%\affil[1]{University of Salzburg}
%\affil[2]{University of Strathclyde}
%\affil[3]{European Central Bank}
\author{Florian Huber\thanks{Corresponding author. Email: \href{mailto: florian.huber@sbg.ac.at}{florian.huber@sbg.ac.at}. Any views expressed in this paper represent those of the authors only and not necessarily  of the ECB or the Eurosystem.   We would like to thank two anonymous referees, Niko Hauzenberger, Michael Pfarrhofer and the participants of the 2019 NBP Workshop on Forecasting, the annual meeting of the Austrian Economic Association (NOeG), the 13$^{th}$ RCEA Bayesian Econometrics Workshop and internal research seminars at the Erasmus University Rotterdam and the University of Salzburg for helpful comments that significantly improved the paper. The first author acknowledges funding from the Austrian Science Fund (FWF): ZK 35. This paper, subject to minor editorial changes, is forthcoming in the Journal of Business \& Economic Statistics.}\\
\emph{University of Salzburg} \and Gary Koop\thanks{Email: \href{mailto:gary.koop@strath.ac.uk}{gary.koop@strath.ac.uk}} \\
\emph{University of Strathclyde} \and Luca Onorante\thanks{Email: \href{mailto:luca.Onorante@ecb.int}{Luca.Onorante@ecb.int}} \\
\emph{European Central Bank}}
\date{}

\maketitle

\date{}

\begin{abstract}
  \noindent Time-varying parameter (TVP) models have the potential to be over-parameterized, particularly when the number of variables in the model is large. Global-local priors are increasingly used to induce shrinkage in such models. But the estimates produced by these priors can still have appreciable uncertainty. Sparsification has the potential to reduce this uncertainty and improve forecasts. In this paper, we develop computationally simple methods which both shrink and sparsify TVP models. In a simulated data exercise we show the benefits of our shrink-then-sparsify approach in a variety of sparse and dense TVP regressions. In a macroeconomic forecasting exercise, we find our approach to substantially improve forecast performance relative to shrinkage alone. 
\end{abstract}

\bigskip
\begin{tabular}{p{0.2\hsize}p{0.65\hsize}} %0.15
\textbf{Keywords:}  & Sparsity, shrinkage, hierarchical priors, time varying parameter regression\\
\textbf{JEL Codes:} & C11, C30, E30, D31 \\
\end{tabular}

\newpage\section{Introduction}\label{sec:intro}
\doublespace
Time-varying parameter (TVP) regressions and Vector Autoregressions (VARs) have enjoyed great popularity among econometricians in recent years as a way of modelling the parameter change that occurs in many macroeconomic and financial time series variables. These are state space models which have been found to work well in forecasting \citep[e.g.][]{giannone2013} and been successfully used for structural economic analysis in a changing environment  \citep[e.g.][]{cogley2005drifts, primiceri2005}. They are flexible and capable of modelling almost any nonlinear relationship between explanatory and dependent variables. 
However, this flexibility comes at a cost: TVP models can be over-parameterized and suffer from the  curse of dimensionality, particularly when the number of potential explanatory variables is large. This can lead to very good in-sample fit, but poor out-of-sample forecast performance. 

There is a large and growing literature that proposes various methods for overcoming these over-parameterization concerns using Bayesian methods  \citep[see, among others,][]{fs_wagner, bkk, kg2014, kowal, uribe_lopes, rockova, koopkorobilisvb, bitto_fs, huberetalJAE, eisenstat_TVPVAR}. These papers propose different approaches to obtain more precise inference. Much of this literature uses hierarchical global-local shrinkage priors. A key property of these priors is that they ensure shrinkage in the sense that they pull coefficients towards zero. However, they do not impose them to be exactly zero and, thus, estimation uncertainty remains. In contrast to shrinkage approaches, selection approaches seek to choose a single sparse specification. That is, they select a particular set of explanatory variables and, by doing so, impose coefficients on non-selected explanatory variables to be zero.\footnote{In the Bayesian literature, there are some global-local priors, such as the spike and slab prior, which do select variables, but these are less popular since Markov Chain Monte Carlo (MCMC) algorithms tend to mix poorly.}   

Which is better: shrinkage or sparsity? The answer to this question depends on the empirical application. In macroeconomics, there is evidence that shrinkage and sparsity can both play a role. For instance, in the case of constant coefficient regressions and VARs, there is debate among Bayesian econometricians as to whether models are sparse (in which case sparsification methods are appropriate) or dense (in which case shrinkage is appropriate). A recent paper, \cite{giannone2017economic}, considers a range of data sets in macroeconomics, microeconomics and finance and finds evidence mostly in favor of dense models, a finding reinforced by \cite{chp}. But there are exceptions to this pattern where sparse models do better. But, instead of choosing one of sparsity or shrinkage, why not do both? This is exactly what recent papers such as \cite{hahncarvalho2015dss} propose. That is, first shrinkage is done using a Bayesian global-local shrinkage prior and then sparsification is done on the resulting estimates. Such an approach could add the benefits of sparsity (i.e. the reduction in estimation error that is important for improving forecasts) along with the benefits of shrinkage which are so useful with dense data sets. {Recent contributions in finance provide  evidence that this works well if interest centers on non-linear modeling of expected returns of companies \citep{fisher2018monotonic} or constructing optimal portfolios \citep{puelz2019portfolio}.}

One consideration that arises in some approaches is computation. Bayesian inference with hierarchical shrinkage priors requires computationally-burdensome MCMC methods. Adding a second sparsification step can greatly increase the burden if this step uses cross-validation methods for choosing key tuning parameters. However, in a recent contribution, \cite{bhattacharya2018signal} propose a  simple algorithm, the signal adaptive variable selector (SAVS), for doing the sparsification step. This involves no tuning parameters and is computationally trivial. \cite{bhattacharya2018signal} provides a theoretical justification for SAVS and shows it to have good empirical performance in simulated and real data contexts. 

The papers cited in the preceding paragraph all relate to constant coefficient regression or VAR models rather than the TVP state space models which are the focus of this paper. We develop Bayesian methods for inference and forecasting in TVP regressions and TVP-VARs which both shrink and sparsify. The shrinkage step can be done using any of the hierarchical shrinkage priors that have been used with TVP regressions. In this paper, we use the Dirichlet-Laplace prior  \citep[see][]{bhattacharya2015dirichlet}, a fully hierarchical variant of the stochastic search variable selection prior \citep[see][]{george1993variable, ishwaran2005spike, george2008bayesian}, the Horseshoe  \citep[see][]{carvalho2010horseshoe}, the Bayesian Lasso of \cite{park2008bayesian} and the Normal-Gamma prior of \cite{griffin2010inference}. The sparsity step is done using the SAVS method of \cite{bhattacharya2018signal}.

Another extension we make in this paper relative to the shrink-then-sparsify methods of \cite{hahncarvalho2015dss} and \cite{bhattacharya2018signal} is that we allow for uncertainty in the sparsified estimates. That is, \cite{hahncarvalho2015dss} and \cite{bhattacharya2018signal} take the posterior mean from the shrinkage step and use only this in the sparsification step. We sparsify every MCMC draw in the shrinkage step, thus allowing for parameter uncertainty. This feature is crucial if interest centers on computing non-linear functions of the parameters (such as higher order predictive distributions) and allows for uncertainty quantification with respect to the chosen model. Our methods are illustrated with simulated and real data and we find them to improve estimation accuracy and forecast performance.

The remainder of this paper is organized as follows: Section \ref{sec:regress} discusses various global-local shrinkage priors in the context of the regression model with constant coefficients. It describes how the sparsification strategy of \cite{bhattacharya2018signal} works in the regression model. Section \ref{ss_tvp} extends these methods to TVP regressions and TVP-VARs. Section \ref{artdat} investigates the performance of our methods relative to non-sparsified alternatives using simulated data from a range of sparse and dense TVP regressions. Section \ref{forecast} carries out a forecasting exercise using TVP-VARs. A comparison of forecasts which are both shrunk and sparsified to those which are only shrunk shows the benefits of doing both. Section \ref{concl} concludes the paper and a technical appendix provides further details on the specific prior setup and the posterior simulation algorithms.

\section{Shrinkage and Sparsification in Regression Models}\label{sec:regress}
In this section we describe the shrinkage and sparsification methods for regression which we build on in this paper. In the next section, we will show how they can be adapted for dynamic regressions and multiple equation models such as VARs. Consider the regression model:
\begin{equation}
y_t = \bm{\beta}^{\prime} \bm{X}_{t} + \varepsilon_t,
\label{reg}
\end{equation}
for $t=1,\ldots,T$ where $y_{t}$ is a scalar dependent variable, $\bm{X}_{t} = (X_{1t}, \dots, X_{Kt})'$ is a $K\times 1$ vector that stacks the explanatory variables $X_{jt}~(j=1,\dots,K)$ and $\bm \beta$ is a $K$-dimensional vector of regression coefficients. The errors are  assumed to be independent and follow a zero mean Gaussian distribution with variance $\sigma^2_\varepsilon$.

When $K$ is large relative to $T$, Bayesians increasingly use hierarchical priors so as to induce shrinkage. Global-local shrinkage priors are particularly popular \citep[see, e.g.,][]{polson2010shrink}. These contain shrinkage that is both global (i.e. common to all parameters) and local (i.e. specific to each parameter). We consider priors which can be represented as scale mixtures of Gaussians. In particular, for the $j^{th}$ regression coefficient we assume: 
\begin{equation}
\beta_j \sim \mathcal{N}(0, \phi_j \lambda), \quad \phi_j \sim f, \quad \lambda \sim g. \label{eq: GL_general}
\end{equation}
Global shrinkage is controlled by $\lambda$ while $\phi_j$ handles the shrinkage of coefficient $j$.
$f$ and $g$ are mixing densities and many different choices have been proposed for them. In this paper, we consider the Horseshoe (HS) prior of \cite{carvalho2010horseshoe}, the Bayesian Lasso (Lasso) of \cite{park2008bayesian}, the Normal-Gamma (NG) prior of \cite{griffin2010inference}, the Dirichlet-Laplace (DL) prior of \cite{bhattacharya2015dirichlet} and the Normal- mixture of Inverse Gamma (NMIG) prior of \cite{ishwaran2005spike}, which is a variant of the stochastic search variable selection (SSVS) prior of \cite{george1993variable, george1997approaches}. All of these are global-local shrinkage priors and differ from one another only in the choices of $f$ and $g$. In addition, and unless  otherwise noted, we use a weakly informative inverted Gamma prior  on $\sigma_\varepsilon^2$ with hyperparameters $d_{\sigma}=e_{\sigma}=0.01$.

Using any of these global-local shrinkage priors, MCMC methods can be used to carry out posterior inference and calculate the posterior mean, $\hat{\bm{\beta}}$. This estimate has been shrunk, but not sparsified. { It could be that many elements of $\hat{\bm{\beta}}$ will be close to zero and thus imply a small but negligible effect of $X_{jt}$ on $y_t$. For large $K$, this potentially leads to overfitting issues which is a direct consequence of the fact that shrinkage is limited by a lower bound on the degree of certainty (since we always have prior scaling parameters that might be close to but not exactly equal to zero).} Sparsification solves this by taking $\hat{\bm{\beta}}$ and setting small elements in it to zero.

{Sparsification has been advocated both as a way of improving model interpretability as well as improving forecasts. In regression models with many explanatory variables, various sparsification approaches have been proposed to select the most important variables so as to simplify the task of interpreting the results \citep[see, e.g.,][]{woody2019}. The influential paper of \cite{barbieriberger} shows that, under certain conditions, the median probability model (i.e. a sparsified model which discards all coefficients with inclusion probabilities below $0.5$) has the best forecast performance. In this paper, we implement sparsification using a recently proposed method where the choice of thresholds is made in an optimal way based on a particular decision theoretic problem.}

We implement sparsification using methods developed in \cite{hahncarvalho2015dss} and \cite{bhattacharya2018signal}. We first define the SAVS estimate and then offer some explanation and motivation for it. 
The SAVS estimate is
\begin{equation}
\overline{\gamma}_j = \text{sign}(\hat{\beta}_j)~ ||\bm{X}_j||^{-2} \text{max}\left( |\hat{\beta}_j|  ~  ||\bm{X}_j||^2 - \kappa_j,~0\right), \label{eq: SAVS}
\end{equation}
with $\bm{X}_j=(X_{j1}, \dots, X_{jT})'$  denoting the $j^{th}$ column of a $T \times K$ matrix  $\bm X=(\bm X_1, \dots, \bm X_T)'$, $(x)_+=\text{max}(x, 0)$ and $\text{sign}(x)$ returns the sign of $x$. Note that this is a soft-thresholding approach where all values of $\overline{\gamma}_j $ below a certain value are set to zero and that it only acts on the posterior mean. 

The sparsified estimate depends on tuning parameters, $\kappa_j$, which determine the thresholds for each coefficient. Various approaches to selecting these have been proposed in the literature including computationally-intensive approaches such as cross-validation. However, \cite{bhattacharya2018signal} come up with a surprisingly simple solution. This is to set:
\begin{equation}
\kappa_j = \frac{1}{|\hat{\beta}_j|^2}. \label{eq: kappa}
\end{equation}
This choice implies a penalty for the $j^{th}$ variable which is ranked in inverse-squared order relative to the magnitude of the $j^{th}$ coefficient. With this choice of thresholds, the SAVS estimate is trivial to calculate.

{To provide some motivation for the SAVS estimate note that (\ref{eq: SAVS}) can be obtained by first solving an optimization problem closely related  to the adaptive Lasso \citep[see][]{zou2006adaptive}:}
\begin{equation}
\overline{\bm \gamma}  =  \underset{ \bm{\gamma}}{\text{arg min}}\left\lbrace \frac{1}{2}\left\| \bm{X}  \hat{\bm \beta} - \bm{X} \bm \gamma \right\|^2_2 + \sum_{j = 1}^{K} \kappa_j |{\gamma}_j|\right\rbrace.\label{eq: loss_vec}
\end{equation}
Equation (\ref{eq: loss_vec}) tries to find a sparse coefficient vector  $\bm \gamma$ that is close to $\hat{\bm \beta}$ while introducing a penalty in case of non-zero elements in $\bm \gamma$. 

The typical way to solve this optimization problem is using a coordinate descent algorithm \citep{friedman2007pathwise}. 
But, as shown in \cite{bhattacharya2018signal}, if you initialize this algorithm at $\hat{\bm \beta}$ and then do one iteration you get precisely the simple algorithm described in (\ref{eq: SAVS}) and (\ref{eq: kappa}). 
It is also noted in \cite{bhattacharya2018signal} that convergence almost always occurs after one iteration and, hence, stopping after one iteration is a sensible thing to do.

One key shortcoming of computing the SAVS estimate is that uncertainty quantification about $\overline{\bm \gamma}$ is not possible and computing non-linear functions of $\overline{\bm \gamma}$ calls for Monte Carlo integration techniques. \cite{bhattacharya2018signal} highlight that one potential solution to this issue is to replace $\hat{\bm{\beta}}$ with a draw from the full conditional posterior distribution of $\bm \beta$. This is an insight we build upon in the context of the TVP models which are the focus of this paper. 

{A recent paper, \cite{woody2019}, develops methods for improving estimates of posterior uncertainty in sparsified regression and nonparametric regression models (but not the TVP state space models). This paper provides further theoretical justification for our approach. It sets up an optimization problem where the goal is to find a parsimonious summary of the posterior which minimizes a loss function which combines model fit with a reward for parsimony or a restriction that the posterior summary lies in a more parsimonious class of models. As a simple example consider a regression with large $K$. A loss function could be chosen which would find the optimal regression model with $p<K$ explanatory variables. In the context of Bayesian MCMC estimation of a regression model, the algorithm of \cite{woody2019} would perform MCMC on the model with $K$ regressors and project each draw into the sparse posterior for the optimal model with $p$ explanatory variables. This is  essentially the same strategy as we will adopt below, although our approach is slightly more general in that our posterior summaries are based on all sparsified MCMC draws. To make this point clear in the context of our simple example, the \cite{woody2019} algorithm would choose one specific optimal set of $p$ explanatory variables \citep[e.g. using the methods of][]{hahncarvalho2015dss} and then project the MCMC draws from the $K$ variable regression into the regression model with the chosen set of $p$ variables. Our algorithm, if used in this simple example, would allow for uncertainty about which specific set of $p$ variables is optimal and, thus, allow for model uncertainty. Apart from this difference, the derivations in \cite{woody2019} provide a theoretical justification for our approach and, in particular, the measures of posterior uncertainty it produces.}

\section{Shrinkage and Sparsification in TVP Models}\label{ss_tvp}

In this section, we develop methods for shrinkage and sparsification in state space models such as the TVP regression and the TVP-VAR. This is achieved using  the non-centered parameterization of \cite{fs_wagner}. We emphasize that the algorithms below do the sparsification at each draw from the MCMC algorithm,  allowing for treatment of uncertainty in the shrinkage step. Thus, the algorithms are averaging over different sparsified estimators in a manner similar to Bayesian model averaging. 

\subsection{The TVP Regression Model}

The TVP regression model used in this paper takes the form:
\begin{align*}
y_t &= \bm{\beta}'_t \bm{X}_t + \varepsilon_t, \\
\bm{\beta}_t &=\bm{\beta}_{t-1}+\bm{w}_t, 
\end{align*}
where all definitions are the same as in (\ref{reg}) except that $\bm{\beta}_t= ( \beta_{1t}, \dots, \beta_{Kt})'$ are dynamic (time-varying) regression coefficients  which follow a random walk with  
 $\bm{w}_t$ being Gaussian innovations with zero mean and variance-covariance matrix $\bm{V}=\text{diag}(v_{1}, \dots, v_{K})$. Each $v_j ~(j=1,\dots, K)$ is a process innovation variance associated with the $j^{th}$ coefficient and thus controls the amount of time-variation in $\beta_{jt}$. 
 
The non-centered parameterization of this model is given by:
\begin{align*}
y_t =& \bm{\beta}_0' \bm{X}_t + \tilde{\bm{\beta}}'_t \sqrt{\bm{V}} \bm{X}_t + \varepsilon_t, \label{eq: NCP_OBS}\\
\tilde{\bm{\beta}}_t =& \tilde{\bm{\beta}}_{t-1} + \bm{\eta}_t,\quad \bm{\eta}_t \sim \mathcal{N}(\bm{0}_K, \bm{I}_K).
\end{align*}
with  the $j^{th}$ element of $\tilde{\bm{\beta}}_t$  given by $\tilde{\beta}_{jt}=\frac{\beta_{jt}-\beta_{j0}}{ \sqrt{v_{j}}}$, $\sqrt{\bm{V}} = \text{diag}(\sqrt{v_1}, \dots, \sqrt{v_K})$ and $\tilde{\bm \beta}_0 =\bm 0_K$.  This equation can be written as:
\begin{equation}
y_t = \bm{\alpha}' \bm{Z}_t + \varepsilon_t,
\label{eq: tvp}
\end{equation}
whereby $\bm{\alpha}  = (\bm{\beta}_0', \sqrt{v_1}, \dots, \sqrt{v_K})'$, $\bm{Z}_t = [\bm{X}'_t,  (  \tilde{\bm{\beta}}_t \odot \bm{X}_t)']'$ and $\odot$ denotes element-wise multiplication. Conditional on knowing the full history of the states in $\bm \tilde{\bm \beta}_t$, (\ref{eq: tvp}) resembles a standard regression model with a (partially) latent covariate vector $\bm Z_t$.

Well-developed MCMC methods exist to carry out Bayesian posterior and predictive inference in state space models such as the TVP regression model under various priors. In this paper, we simulate the full history of the normalized dynamic regression coefficients $\{\tilde{\bm \beta}_t\}_{t=1}^T$ using the forward-filtering backward-sampling algorithm proposed in \cite{carterkohn} and \cite{fs1994}. Conditional on $\tilde{\bm \beta}_t$, (\ref{eq: tvp}) is a standard regression model, implying that we can simulate $\bm \alpha$ from a Gaussian full conditional posterior distribution and $\sigma^2_\varepsilon$ from an inverted Gamma distribution. The corresponding moments take standard forms and are presented in Appendix \ref{sec: MCMC}.

We propose to do shrinkage on $\bm{\alpha}$ using the global-local mixture priors mentioned in the previous section and  described in Appendix \ref{sec: GL_priors}. That is, conditional on a draw of the full history of the states, $\{\tilde{\bm{\beta}}_t\}_{t=1}^{T}$, we have the regression model given in (\ref{eq: tvp}), and shrinkage can be done exactly as described in the preceding section.  For each of the global-local mixture priors we consider, MCMC methods for drawing $\bm{\alpha}$ and $\sigma^2_\varepsilon$, conditional on draws of the states exist. For the Dirichlet-Laplace prior we follow the methods of \cite{bhattacharya2015dirichlet}. For the NMIG specification, we adopt the algorithm proposed in \cite{ishwaran2005spike} while  for the Horseshoe, the MCMC algorithm developed in \cite{makalic2016simple} is used. Since the Normal-Gamma prior nests the Bayesian Lasso, we adopt the algorithm put forth in \cite{griffin2010inference} (see Appendix  \ref{sec: GL_priors} for further details). 

As highlighted in Section \ref{sec:regress}, using shrinkage implies that elements in $\bm \alpha$ are pushed to zero and elements in $\bm Z_t$ might have a small effect on $y_t$. However, in the TVP regression setting, this problem is intensified since the state equation can be written in terms of the sum of the past shocks to the states $\bm w_t$. The corresponding variance of $\bm \beta_t$ thus increases with time and values of $\sqrt{v_j}$ that are close to zero could still induce large aggregate movements in $\beta_{jt}$ over time. In such a situation, sparsification might help since setting $\sqrt{v_j}=0$ directly implies that $\beta_{jt}=\beta_{jt-1}$ for all $t$.

Given a draw from the posterior of $\bm \alpha$, denoted as $\bm \alpha^{(n)}$,  from any of the MCMC algorithms is sparsified using SAVS.  Applying the SAVS estimator in (\ref{eq: SAVS}) to each draw from the posterior of $\bm \alpha$ yields:
\begin{equation}
\overline{\gamma_j}^{(n)} = \text{sign}\left(\alpha^{(n)}_j\right)~ ||\bm{Z}_j||^{-2} \left( |{\alpha}^{(n)}_j|  ~  ||\bm{Z}_j||^2 - \kappa_j\right)_+, \text{ for } n=1,\dots,N \label{eq:SAVSSS}
\end{equation}
where $\bm Z_j$ denotes the $j^{th}$ column of $\bm Z = (\bm Z_1, \dots, \bm Z_T)'$, $\kappa_j = |{\alpha}^{(n)}_j|^{-2}$ and $N$ denotes the number of post burn-in MCMC draws. This procedure effectively allows for uncertainty quantification and the computation of potentially non-linear functions of the sparsified parameters such as higher-order forecasts or impulse response functions.  Thus, one can think of our proposed procedure as an approximate MCMC algorithm which draws from the sparsified conditional posterior  
$p(\overline{\bm{\gamma}}| \bm \alpha, \bm Z)$.\footnote{The algorithm is approximate since $\sigma^2_\varepsilon$ does not play a role in the SAVS algorithm. If desired, after each sparsification, one could take a draw of $\sigma_\varepsilon^2$ conditional on the sparsified estimates.} Hence, forecasts produced will average over different sparsified models. That is, one MCMC draw will lead to one particular sparsified model which is used for forecasting, another draw may choose another sparsified model to produce forecasts. Hence, what we are proposing is similar in spirit to Bayesian model averaging.  {This feature allows us to calculate posterior inclusion probabilities (PIPs) for each variable. The PIP for a given coefficient is the proportion of MCMC draws for which the coefficient is not set to zero.}

Another possibility would be to use the SAVS algorithm directly on the posterior mean of $\bm{\alpha}$ as is done by \cite{hahncarvalho2015dss} and \cite{bhattacharya2018signal}. This procedure yields a point estimate for the time-invariant coefficients and the state innovation variances. {However, one shortcoming of doing this is that $\bm Z_t$ includes latent quantities that need to be integrated out or a plug-in estimate (such as the posterior mean) might be used. However, as \cite{puelz2017variable} note, this could negatively impact inference since the corresponding uncertainty surrounding $\bm Z_t$ is ignored. Our approach circumvents this by integrating out the latent states contained in $\bm Z_t$. In addition, if the researcher wishes to select a single sparse model, as produced by sparsifying the posterior mean directly, our approach provides an alternate way of choosing the sparsity pattern based on PIPs. }

%This would be similar in spirit to a Bayesian model selection strategy where a single sparsified model was chosen for forecasting. But this would ignore model uncertainty and, in addition, would be problematic since $\bm Z_t$ is partially latent. 

Another point worth emphasizing about our algorithm is that it is fast. Relative to the computational time required to do MCMC, adding the SAVS step increases the computational burden by a trivial amount. For any empirical specification where MCMC is possible, our proposed algorithm is also possible. Of course, if $K$ is too large, then MCMC methods may be computationally infeasible. In such a case, variational Bayesian methods may be a practical alternative  \citep[see][]{koopkorobilisvb}. But with variational Bayes methods, the SAVS algorithm would be applied on the approximate posterior mean and model uncertainty ignored.\footnote{It would be possible to surmount this drawback of variational Bayes by first using variational Bayes to obtain an approximation to the posterior and then applying the SAVS algorithm to draws from this approximation. But this would be computationally demanding, thus undermining the main advantage of variational Bayes.}

\subsection{The TVP-VAR}\label{sec:tvpvar}

The shrink-then-sparsify algorithm we propose for the TVP regression can be extended to handle the TVP-VAR in a straightforward fashion. The idea is to transform the TVP-VAR so that the error covariance matrix in the measurement equation is diagonal. Then the TVP regression algorithm of the preceding sub-section can be applied one equation at a time. Equation-by-equation estimation of VARs is done in several recent papers using transformations similar to the one used here \citep[see, e.g.,][]{kastnerhuber2017, kpp2019, ccm2016} and the reader is referred to these papers for further details about the computational advantages of this approach. With macroeconomic data it is often important to add stochastic volatility (SV), which leads us to the TVP-VAR-SV specification described in this section.  

Let $\bm{y}_{t}$ be an $M \times1$ vector of endogenous variables for $t=1,\ldots,T$. The TVP-VAR-SV can be written as:
\begin{equation}
\bm{y}_{t}= (\bm{I}_M \otimes \bm{X}_t) \bm{\beta}_t + \bm  \varepsilon_t,
\label{VAR}
\end{equation}
where
$\bm{X}_t = (\bm{y'}_{t-1}, \dots, \bm{y'}_{t-P}, 1)'$ contains the $P$ lags of $\bm{y}_t$ and an intercept, $\bm{\beta}_t$ is  the vector $K=M (MP +1)$ coefficients at time $t$ which is assumed to evolve according to a multivariate random walk. 
The errors are independent over time with $\bm \varepsilon_t \sim \mathcal{N}(\bm{0}_M, \bm{\Sigma}_t)$.
 $\bm{\Sigma}_t$ is the time-varying error covariance matrix with
\begin{equation*}
\bm{\Sigma}_t = \bm{U}_t \bm{H}_t \bm{U}_t'.
\end{equation*}
Let $\bm{U}_t$ denote a lower uni-triangular matrix and $\bm{H}_t = \text{diag}(e^{h_{1t}}, \dots, e^{h_{Mt}})$. The $M (M-1)/2$ free elements in $\bm U_t$ follow independent random walks while the $h_{jt}$'s are log-volatilities that evolve according to AR(1) processes,
\begin{equation}
h_{jt} = \mu_j +  \rho_j (h_{jt-1}-\mu_j) + \eta_t, \quad \eta_t \sim \mathcal{N}(0, \sigma_{\eta, j}^2). \label{eq: stateh}
\end{equation}
Here, we let $\mu_j$ denote the unconditional mean, $\rho_j$ the persistence parameter and $\sigma_{\eta, j}^2$ the error variance of the log-volatility process.  The initial state $h_0$ is drawn from the stationary distribution of the process. The prior specification on the parameters of the log-volatility equation closely follows \cite{kastner2014ancillarity}. Specifically, we use a zero mean Gaussian prior with variance $10^2$ on $\mu_j$, a Beta prior on $\frac{\rho_j+1}{2} \sim \mathcal{B}(25, 5)$ and a Gamma prior on $\sigma_{\eta, j}^2 \sim \mathcal{G}(1/2, 1/2)$. This Gamma prior translates into a Gaussian prior on $\pm \sigma_{\eta, j}$ with zero mean and unit variance.  In the MCMC algorithm, the full history of $h_{jt}$ as well as the parameters of  equation (\ref{eq: stateh}) are obtained  using the algorithm proposed in \cite{kastner2014ancillarity}. This algorithm exploits the centered and non-centered parameterization of the non-linear state space model to increase sampling efficiency and samples the full history of the log-volatilities from a $(T-1)$-dimensional multivariate Gaussian distribution.

As noted in \cite{ccm2016, kastnerhuber2017, kpp2019}, computation is greatly simplified if the model is transformed so that the errors in different equations are independent of one another. This can  be achieved by augmenting the $i^{th}$ equation in the system with the contemporaneous values of the first $i-1$ elements in $\bm y_t$. That is, if $y_{it}$ is the $i^{th}$ variable (for $i>1$), we can write the TVP-VAR-SV as a set of $M$ unrelated TVP regressions:
\begin{equation}
y_{it} = \bm{X}'_t  \bm{\beta}_{it} + \sum_{j=1}^{i-1} u_{ij,t} y_{jt}+\eta_{it},\quad  \eta_{it}\sim \mathcal{N}(0, e^{h_{it}}),
\label{eqeq}
\end{equation}
where $\eta_{it}$ and $\eta_{jt}$ are independent for $i\ne j$, $\bm{\beta}_{it}$ denotes the elements of $\bm{\beta}_t$ in the $i^{th}$ equation and $u_{ij,t}$ are the elements of $\bm{U}_t^{-1}$ for $i=2,\dots, M; j=1,\dots,i-1$.

We then write the TVP-VAR-SV using the non-centered parameterization. For equation $i$ we obtain:
\begin{equation}
y_{it} = \bm{X}'_t  \bm{\beta}_{i0} +(\sqrt{\bm{V_{i}^{\beta}}} \bm{X}_t)'  \tilde{\bm{\beta}}_{it} + \sum_{j=1}^{i-1} u_{ij,0} y_{jt}+ \sum_{j=1}^{i-1} \sqrt{v_{ij}^{u}} \tilde{u}_{ij,t} y_{jt}+\eta_{it},\quad  \eta_{it}\sim \mathcal{N}(0, e^{h_{it}}).
\label{eqeq: TVPVAR_noncen}
\end{equation}
Here, we let $\sqrt{\bm{V_{i}^{\beta}}} = \text{diag}\left(\sqrt{v_{i1}^{\beta}}, \dots, \sqrt{v_{iK}^{\beta}}\right)$ and $\sqrt{v_{ij}^{u}}$ denotes the standard deviation of the error in the random walk state equation for the $j^{th}$ VAR coefficient in the $i^{th}$ equation. Similarly,  $\sqrt{v_{ij}^{u}}$ is the standard deviation for the random walk state equation for the elements of $\bm{U}_t$.  Thus, $\tilde{\bm{\beta}}_{it}$ and $\tilde{u}_{ij,t}$ are the states for equation $i$ and the  shocks in the corresponding state equations have unit standard deviation.

Since the errors in the different equations are independent of one another, estimation of  one equation at a time using the algorithm of the preceding sub-section, including the SAVS step detailed in (\ref{eq:SAVSSS}), can be done. Computation is also sped up since parallelization is feasible. Note also that, since the coefficients in $\bm{U}_t^{-1}$ are appearing as regression coefficients in (\ref{eqeq}), these can also be shrunk and sparsified.   In large TVP-VARs, where there are many such error covariance terms, this is potentially beneficial for forecasting purposes. Notice that we do not only obtain a sparse  error covariance matrix but also allow for checking whether the corresponding free elements are time-varying or constant.

\section{Evidence Using Artificial Data}\label{artdat}
In this section, we present evidence on the performance of the proposed methodology using artificial data generated from different TVP regression models. 
Across the different data generating processes (DGPs),  the covariates are drawn from a Uniform distribution bounded between $-1$ and $1$.  The $\bm \beta_t$'s are generated using the non-centered parameterization with $\bm \beta_0 \sim \mathcal{N}(\bm 0_K, 0.1^2  \bm I_K )$ and $\pm \sqrt{v_j} \sim \mathcal{N}(0, 0.1^2), j=1,\dots, K,$ while differing percentages of the elements in $\bm \alpha$ are randomly set to zero. The measurement error variance $\sigma^2_\varepsilon$ is set equal to $0.1^2$. 

\begin{figure}[t]
\caption{Sparsified and non-sparsified posteriors of $\beta_{jt}$ for a large time-varying parameter model}\label{fig_ss}
\begin{minipage}{.33\linewidth}
\vspace{0.2cm}
\centering{(a) DGP with $\beta_{jt}=0$ for all $t$}
\end{minipage}
\begin{minipage}{.33\linewidth}\vspace{0.2cm}

\centering{(b) DGP with $\beta_{jt}=\beta_{jt-1}$ for all $t$}
\end{minipage}
\begin{minipage}{.33\linewidth}\vspace{0.2cm}

\centering{(c) DGP with $\beta_{jt}$ time-varying}
\end{minipage}\\
\begin{minipage}{.33\linewidth}
\includegraphics[scale=0.33,page=1]{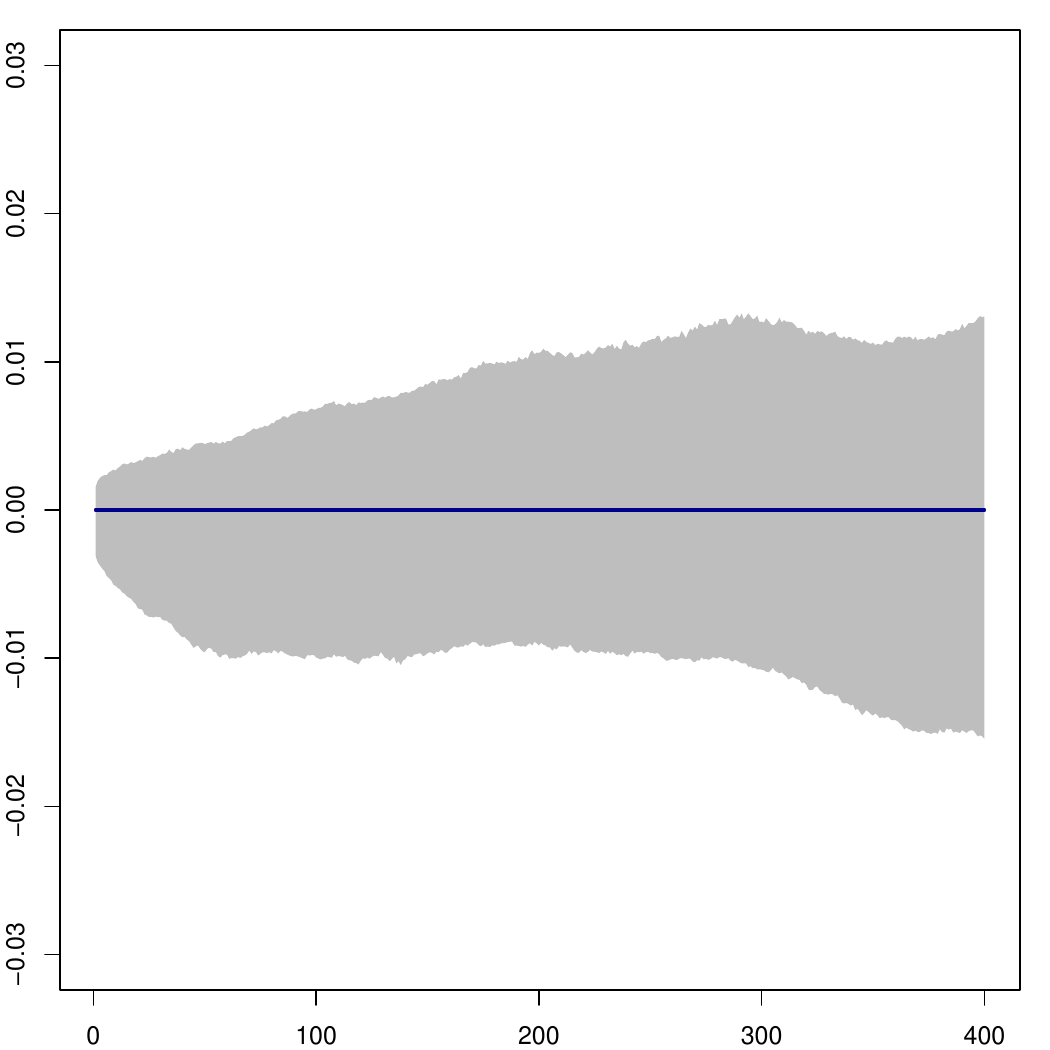}
\end{minipage}
\begin{minipage}{.33\linewidth}
\includegraphics[scale=0.33,page=23]{Results/parms.pdf}
\end{minipage}
\begin{minipage}{.33\linewidth}
\includegraphics[scale=0.33,page=14]{Results/parms.pdf}
\end{minipage}\\
\begin{minipage}{\linewidth}
\vspace{.1cm}
\textbf{Notes:} The dark blue line denotes the true $\beta_{jt}$ over time, the gray shaded area represents 5th and 95th credible sets of the posterior of $\beta_{jt}$ under a HS prior. Purple lines represent 5th and 95th credible sets of the sparsified posterior.
\end{minipage}
\end{figure}

Before presenting results using repeated samples, the main features of sparsification are illustrated in Figure \ref{fig_ss}. The results in the three panels of the figure are based on the Horseshoe prior and use three different single artificial data sets obtained by simulating $T=400$ observations from a large ($K=30$) dynamic regression model.
Figure \ref{fig_ss}(a) plots posterior features of $\beta_{jt}$ against time for a case where it is zero (i.e. the DGP is one where $j^{th}$ regressor is not selected) using a non-sparsified and sparsified estimator. Note that the sparsified estimator is precisely correct, it sets $\beta_{jt}=0$ with probability one. Thus, it exactly coincides with the true value and cannot be seen in Figure \ref{fig_ss}(a). The non-sparsified posterior distribution, although the posterior mean is very close to the correct value, has a credible interval that is non-negligible. This reflects estimation uncertainty and could spill over into poor forecast performance using the non-sparsified posterior. The performance of the SAVS algorithm when $\beta_{jt}$ is a non-zero constant (i.e. the DGP is one where $\beta_{jt}=\beta_{jt-1}$ for all $t$) is shown in Figure \ref{fig_ss}(b). In this case, the posterior distributions of the sparsified and non-sparsified models almost coincide. Notice, however, that the credible sets are constant over time for the sparsified model, indicating that the corresponding element in $\sqrt{\bm V}$ is set equal to zero throughout all iterations of the MCMC algorithm. In contrast, Figure \ref{fig_ss}(c)  illustrates a case where $\beta_{jt}$ is non-zero and time-varying. Notice  that  the sparsified and non-sparsified posterior distributions are close to being identical. In this case, it is not desirable to sparsify the corresponding elements in $\bm \alpha$ and the SAVS algorithm is not doing so. Thus, regardless of whether a coefficient is zero, a non-zero constant or time-varying, this figure indicates that our methods estimate it well. They work better than the non-sparsified alternative in cases where there is sparsity and equally well in non-sparse cases. 

Table \ref{tab:artdat} presents evidence for the importance of sparsification and shrinkage in TVP regression models using different data configurations,  priors,  numbers of regressors and  sample sizes. The DGP described above is modified to  reflect varying degrees of sparsity. These different sparsity levels  are  labeled sparse  (with $90$\% zeros in $\bm \alpha$), moderate (with $70$\% zeros) and dense (with $30$\% zeros). To assess how our techniques perform across model dimensions and length of time series involved,  we consider variants of each sparsity level with $K=5, 15$ and $30$ explanatory variables and $T=250$ and $400$ observations. The latter are typical values in quarterly and monthly macroeconomic data sets. For each DGP, we generate $100$ artificial data sets and then run each through a sparsified and non-sparsified algorithm using each of the five global-local shrinkage priors listed in Section \ref{sec:regress}. We also include a non-informative prior (labelled Flat in the table) which does not do shrinkage. Posterior medians of $\bm \beta_{t}$ are produced and the absolute value of the difference between these and the true value used in the DGP is calculated. The figures in Table \ref{tab:artdat} are averages taken over three dimensions: i) the $100$ simulated data sets, ii) time and iii) the $K$ elements of $\bm \beta_t$.

Table \ref{tab:artdat} shows the value of sparsification, particularly with sparse DGPs. With the latter, mean absolute errors (MAEs) are lower than their non-sparsified counterpart for every prior and choice for $T$ and $K$. But even in moderately dense specifications, sparsification lowers MAEs in most cases. In the dense specification, sparsification does not improve upon the single best performing non-sparsified model specification. However, in that situation, accuracy differences are found to be negligible. 

\begin{table}[h!]
\caption{Mean absolute errors between the true realization of the $\bm \beta_t$s and the posterior median}
\centering
\scalebox{.86}{
\begin{tabular}{lrrrrrrrrrrrr}
\toprule
  &\multicolumn{5}{c}{Non-sparsified} & \multicolumn{5}{c}{Sparsified}\\
 Sparsity level& Flat & DL & Lasso & NG & HS &  NMIG & \hspace{.3cm} Flat & DL & Lasso & NG & HS &  NMIG \\ 
   \midrule
  \multicolumn{13}{c}{\hspace{2.8cm}\textbf{Small} $(K=5)$}\\  
  \multicolumn{13}{c}{\hspace{2.8cm}$T=250$}\\
    \midrule
   Dense 				& 7.97 & 7.51 & 8.08 & 7.33 & 7.65 & 7.91 & 7.85 & 7.91 & 8.33 & 7.48 & 7.72 & 8.28 \\             
  Moderate 					& 4.66 & 4.59 & 4.47 & 4.97 & 4.27 & 3.62 & 4.06 & 4.53 & 4.29 & 5.11 & 4.30 & 3.75 \\         
  Sparse 					& 3.63 & 3.05 & 3.46 & 2.73 & 3.48 & 3.33 & 2.84 & 2.96 & 3.04 & 2.58 & 3.15 & 3.17 \\         
 
     	 \multicolumn{13}{c}{\hspace{2.8cm}$T=400$}\\  \midrule
   Dense 			& 7.57 & 7.14 & 6.89 & 6.40 & 7.65 & 7.03 & 7.45 & 7.65 & 7.15 & 6.42 & 7.91 & 7.46 \\                                
   Moderate 		& 4.41 & 4.10 & 4.27 & 5.13 & 3.89 & 4.42 & 3.94 & 4.08 & 4.21 & 5.05 & 4.17 & 4.53 \\                               
   Sparse 				& 2.90 & 2.74 & 3.40 & 2.80 & 2.44 & 3.43 & 2.26 & 2.64 & 2.93 & 2.70 & 2.19 & 3.23 \\                            
  \midrule
  \multicolumn{13}{c}{\hspace{2.8cm}\textbf{Medium} $(K=15)$}\\  
  \multicolumn{13}{c}{\hspace{2.8cm}$T=250$}\\
    \midrule
   Dense 				& 11.60 & 10.88 & 9.51 & 9.39 & 8.72 & 9.76 & 11.32 & 11.78 & 9.18 & 9.24 & 8.83 & 10.07 \\                          
  Moderate 				& 3.46 & 3.98 & 5.22 & 3.27 & 3.64 & 3.03 & 2.29 & 3.48 & 4.84 & 2.88 & 3.39 & 2.90 \\                               
  Sparse 				& 3.21 & 1.81 & 2.44 & 2.09 & 1.84 & 2.04 & 2.02 & 1.40 & 1.48 & 1.79 & 1.63 & 1.55 \\                               
 
     	 \multicolumn{13}{c}{\hspace{2.8cm}$T=400$}\\  \midrule
   Dense 				& 9.87 & 10.77 & 9.60 & 8.19 & 8.44 & 8.72 & 9.61 & 11.50 & 9.70 & 8.07 & 8.95 & 9.41 \\           
   Moderate 				& 2.93 & 3.09 & 3.78 & 3.62 & 3.86 & 3.83 & 1.99 & 2.72 & 3.00 & 3.20 & 3.67 & 3.57 \\        
   Sparse 					& 2.30 & 2.14 & 2.37 & 1.75 & 1.27 & 1.90 & 1.23 & 1.89 & 1.96 & 1.51 & 1.08 & 1.39 \\         
  \midrule
  \multicolumn{13}{c}{\hspace{2.8cm}\textbf{Large} $(K=30)$}\\  
  \multicolumn{13}{c}{\hspace{2.8cm}$T=250$}\\
    \midrule
   Dense 							& 15.40 & 14.18 & 14.30 & 12.84 & 13.97 & 12.32 & 15.04 & 15.14 & 13.64 & 12.88 & 14.06 & 12.64 \\                   
   Moderate	 							& 5.24 & 3.48 & 3.84 & 3.26 & 2.44 & 2.38 & 4.27 & 2.83 & 2.74 & 2.64 & 2.04 & 2.11 \\                          
    Sparse 							& 2.53 & 1.33 & 2.48 & 1.36 & 1.72 & 1.67 & 1.17 & 0.79 & 1.59 & 0.97 & 1.59 & 1.07 \\                                \multicolumn{13}{c}{\hspace{2.8cm}$T=400$}\\  \midrule
   Dense 					& 13.71 & 12.48 & 12.29 & 13.50 & 12.39 & 11.68 & 13.33 & 13.27 & 11.83 & 13.22 & 12.17 & 12.06 \\                               
   Moderate 				& 4.73 & 3.43 & 3.71 & 4.06 & 2.64 & 2.27 & 3.73 & 2.79 & 2.69 & 3.92 & 2.38 & 1.76 \\                                          
   Sparse 					& 1.78 & 1.36 & 1.47 & 1.32 & 0.52 & 1.43 & 0.63 & 0.75 & 0.60 & 0.89 & 0.44 & 0.78 \\                                           
   \bottomrule
     \multicolumn{13}{p{17.4cm}}{\footnotesize{\textbf{Notes:} The mean is taken over time, over all parameters and over all artificial data sets. All mean absolute errors are multiplied by $100$. Flat refers to a dynamic regression model with a loosely informative prior, DL to the Dirichlet-Laplace prior, Lasso to the Bayesian Lasso, NG to the Normal-Gamma prior, HS to the Horseshoe, and NMIG to the Normal-mixture of Inverse Gamma prior. }}
\end{tabular}
}
\label{tab:artdat}
\end{table}

The benefits of shrinking and sparsifying increase with the number of explanatory variables. Note, for instance, that the unsparsified Flat prior model does not perform that poorly when $K=3$ and $15$, but displays a weak performance when $K=30$. In fact, when $K=3$, the Flat prior works quite well with the sparse specification, provided sparsification is done. This indicates that there are some cases where sparsification is more important than shrinkage. 

The choice of $T$ has little impact on the results. In a regression model with constant parameters, we would expect sparsification to be  less important as the sample size increases since, with longer time series, the estimation error would decrease. However, with TVP regressions, the number of parameters is also increasing with the sample size which negates this effect. Thus, even with large numbers of observations, the researcher working with TVP models can still benefit from sparsification. 

With regards to the different global-local shrinkage priors, no clear pattern emerges where one performs consistently the best across different specifications. When $K=30$ and the DGP is sparse, DL (for $T=250$) and HS (for $T=400$) models that are sparsified are the best performers. When $K=30$ and the DGP is dense, the accuracy of both, the DL and the HS prior deteriorates slightly while the unsparsified NMIG model shows the best performance. Notice that in this situation, accuracy differences across the sparsified and non-sparsified NMIG specification are, however, quite small.

From this discussion it is apparent that identifying a default prior choice is difficult.  One key take away from this analysis, however, is that if the DGP is sparse,  flexible shrinkage specifications such as the HS, the DL  and the NMIG prior in combination with the SAVS algorithm provide accurate parameter estimates.  Overall, the table tells a story of the importance of both shrinkage and sparsity, especially in large models, with the precise choice of shrinkage prior being of lesser importance.   

In the next step, we assess how well the SAVS algorithm identifies true zeros in $\bm \alpha$. Table \ref{tab:hitrates} shows average hit rates that measure the percentage of correctly estimated zeros using the SAVS algorithm. From this table, we observe that irrespective of the priors used, our approach works well in identifying the true level of sparsity. For sparse situations, the fraction of correctly identified zeros is often above 95\% for most shrinkage priors and model sizes considered. In the case of a Flat prior, we observe values just above 90\%, which is remarkable but still well below the percentages observed for the different shrinkage specifications under scrutiny. This slightly weaker performance can be traced back to the fact that without shrinkage, values in $\bm \alpha$ are not pushed to zero and the corresponding penalty $\kappa_j$ is too small. Consistent with the findings in Table \ref{tab:artdat}, we find no discernible differences in performance across the different shrinkage priors, with all of them displaying a strong performance. In fact, in a sparse setting with $K=30$, the SAVS algorithm identifies almost all zeros correctly, with hit rates being above $99$\%.
\begin{table}[h!]
\caption{Average hit rates of the SAVS algorithm across different prior specifications}
\centering
\scalebox{.86}{
\begin{tabular}{lrrrrrrr}
  \toprule
 Sparsity level & Flat & DL & Lasso & NG & HS &  NMIG  &  \\ 
     \multicolumn{7}{c}{\textbf{Small} $(K=5)$}\\
  \multicolumn{7}{c}{$T=250$}\\
    \midrule
Dense				& 86.50 & 73.60 & 81.30 & 79.90 & 78.70 & 79.60 \\           
  Moderate 			& 88.40 & 93.10 & 92.00 & 94.10 & 91.80 & 91.40 \\           
  Sparse 			& 90.40 & 96.10 & 96.40 & 96.70 & 96.40 & 94.00 \\           
  \multicolumn{7}{c}{$T=400$}\\  \midrule
  Dense 					& 85.60 & 74.80 & 80.20 & 83.10 & 81.50 & 74.90 \\   
  Moderate 					& 89.60 & 92.90 & 93.40 & 94.90 & 93.40 & 91.10 \\   
  Sparse 					& 91.70 & 97.40 & 96.10 & 97.60 & 97.30 & 95.60 \\   
    \midrule
      \multicolumn{7}{c}{\textbf{Medium} $(K=15)$}\\
  \multicolumn{7}{c}{$T=250$}\\
    \midrule
Dense			& 79.17 & 77.63 & 79.37 & 80.03 & 79.07 & 78.23 \\           
  Moderate 			& 83.93 & 97.97 & 95.70 & 97.87 & 96.50 & 96.47 \\       
  Sparse 		& 91.23 & 99.17 & 98.00 & 99.17 & 99.47 & 98.97 \\           
  \multicolumn{7}{c}{$T=400$}\\  \midrule
  Dense 				& 82.57 & 80.40 & 82.27 & 82.97 & 80.73 & 78.33 \\  
  Moderate 				& 87.00 & 98.33 & 96.47 & 97.70 & 96.73 & 96.67 \\  
  Sparse 				& 93.97 & 99.67 & 98.83 & 99.57 & 99.73 & 98.97 \\  
       \midrule
      \multicolumn{7}{c}{\textbf{Large} $(K=30)$}\\
  \multicolumn{7}{c}{$T=250$}\\
    \midrule
Dense			& 68.57 & 72.57 & 76.05 & 76.05 & 76.47 & 75.38 \\               
  Moderate 			& 72.40 & 98.83 & 95.42 & 97.75 & 96.93 & 97.13 \\           
  Sparse 				& 85.12 & 99.70 & 97.73 & 99.62 & 99.62 & 99.55 \\       
  \multicolumn{7}{c}{$T=400$}\\  \midrule
  Dense 					& 71.30 & 78.50 & 78.35 & 79.23 & 77.80 & 76.37 \\ 
  Moderate 				 & 80.68 & 99.17 & 96.32 & 98.57 & 97.17 & 97.38 \\    
  Sparse 				& 91.60 & 99.82 & 98.35 & 99.80 & 99.65 & 99.53 \\     
   \bottomrule
        \multicolumn{7}{p{10.3cm}}{\footnotesize{\textbf{Notes:} The mean is computed over all parameters and  artificial data sets. Flat refers to a dynamic regression model with a loosely informative prior, DL to the Dirichlet-Laplace prior, Lasso to the Bayesian Lasso, NG to the Normal-Gamma prior, HS to the Horseshoe, and NMIG to the Normal-mixture of Inverse Gamma prior.}}
\end{tabular}
}
\label{tab:hitrates}
\end{table}

To sum up, this discussion highlights that sparsification  improves estimation accuracy. These improvements tend to increase with model size and the level of sparsity of the DGP. Among the set of competing shrinkage priors, we find no single best performing specification. In terms of correctly predicting zeros in $\bm \alpha$, we found that SAVS works well across all shrinkage priors considered, often correctly identifying above $99$\% of the zeros. At this point, and before proceeding to the empirical application, it is worth emphasizing that our analysis only considers whether our shrink-then-sparsify approach  improves accuracy of point estimates, ignoring a potential bias-variance tradeoff. One key finding is that applying SAVS never significantly decreases estimation accuracy and correctly predicts a large fraction of true zeros. {In light of Figure \ref{fig_ss}, this indicates that, by zeroing out shrunk coefficients, our approach pushes the posterior variance to zero and this could increase predictive accuracy in forecasting applications.}

%strongly reduces parameter uncertainty.

\section{Forecasting US Macroeconomic Variables}\label{forecast}
In this section we present results from a forecasting exercise using US quarterly macroeconomic data taken from the FRED-QD database \citep[see][]{mccracken2016fred} that span the period from 1959Q1 to 2017Q4. We focus on forecasting GDP, inflation (based on the GDP deflator) and the Fed Funds rate (henceforth labeled focus variables). Table \ref{tab:data-descr1} provides details on the specific variables included alongside transformations used. 

We use the TVP-VAR-SV of Sub-section \ref{sec:tvpvar} combined with the same set of global-local shrinkage priors as in the preceding section. The only specification we do not consider here is the TVP-VAR-SV with a flat prior since this model performs poorly in out-of-sample forecasting and large dimensions.\footnote{The results for the flat prior model are available upon request from the authors.} {More specifically, using weakly informative priors leads to overfitting, which in turn translates into model instability since no  penalty is introduced to rule out explosive regions of the parameter space. \cite{woody2019} note that in such situations, we fit the noise in the first stage, leading to insufficient posterior variability in the summary.}

For each prior, we use non-sparsified and sparsified versions of the model in order to produce the forecasts. We forecast with small ($M=3$), medium ($M=8$) and large ($M=20$) data sets and set the lag length equal to $2$. Thus, the dimension of the state space in the TVP-VAR-SVs ranges from being moderate to huge. Our forecast evaluation begins in 1997Q1 and runs to the end of the sample. We use root mean squared forecast errors (RMSEs) to evaluate the quality of the point forecasts and average log predictive likelihoods (LPLs) to evaluate the quality of predictive densities. Both are benchmarked relative to a VAR-SV with DL prior, a specification that works well for US macroeconomic data \citep[see][]{kastnerhuber2017}.  This is identical to the TVP-VAR-SV with DL prior except that the DL prior now applies directly to the constant VAR coefficients while $\sqrt{\bm V_i^\beta}$ and $\sqrt{v_{ij}^{u}}$ are set equal to zero for all $i, j$. The VAR is transformed to allow for equation-by-equation estimation as described in Sub-section \ref{sec:tvpvar}.  

\begin{figure} \caption{Heatmaps of posterior inclusion probabilities (PIPs) for the three focus variables in $t=2017:Q4$}\vspace{.3cm}\label{fig: heatmap}
\begin{minipage}{\linewidth}
\centering  PIPs for Constant Coefficients ($\beta_{i0}$ and $u_{ij,0}$ in equation \ref{eqeq: TVPVAR_noncen})
\vspace{.3cm}
\end{minipage}\\
\begin{minipage}{.3\linewidth}
\centering   $\bm y_{t-1}$
\end{minipage}
\begin{minipage}{.3\linewidth}
\centering  $\bm y_{t-2}$
\end{minipage}
\begin{minipage}{.3\linewidth}
\centering   Elements of $\bm{U}_{0}$
\end{minipage}\\
\begin{minipage}{1\linewidth}
\includegraphics[scale=.45]{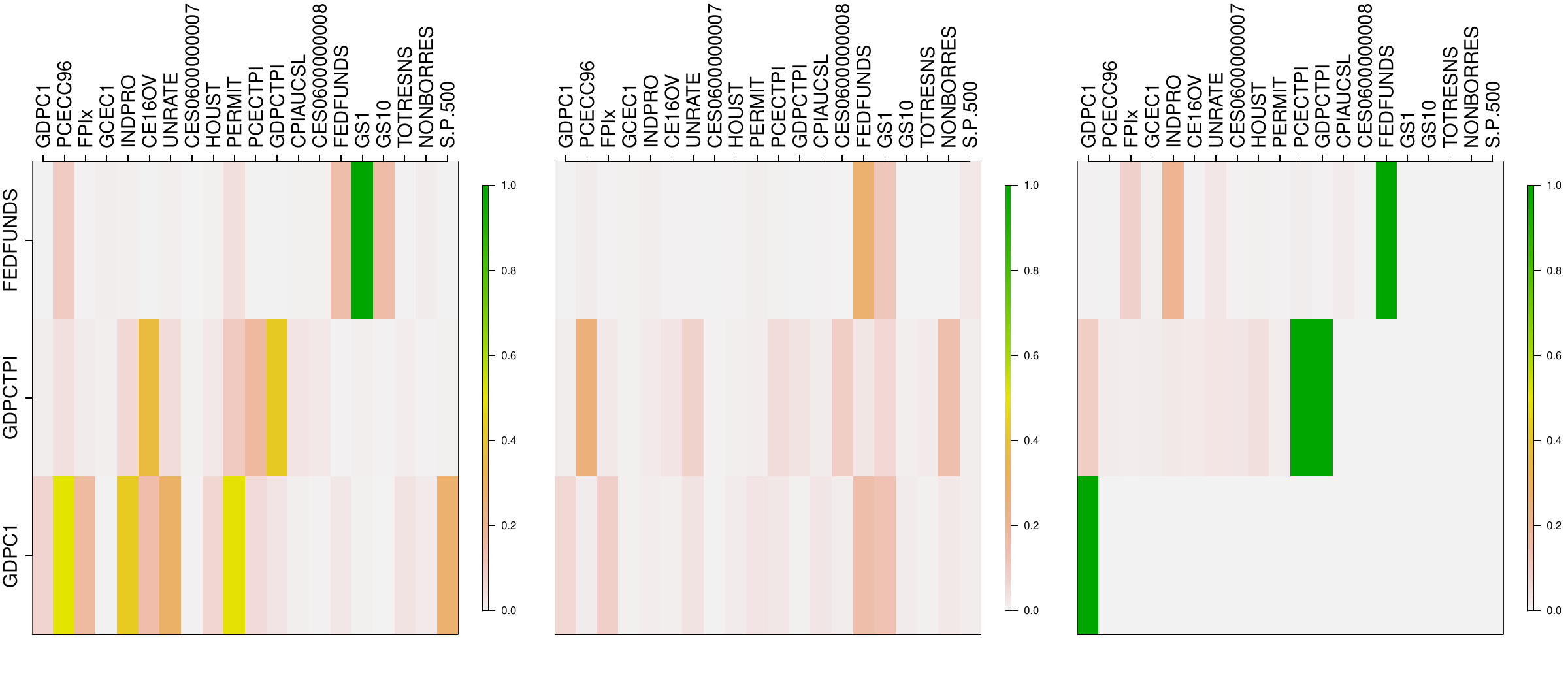}
\end{minipage}\\
\begin{minipage}{\linewidth}
\centering  PIPs for Time-Varying Coefficients ($v_{ij}^{\beta}$ and $v_{ij}^{u}$ in equation \ref{eqeq: TVPVAR_noncen})
\end{minipage}\\
\begin{minipage}{.3\linewidth}
\centering   $\bm y_{t-1}$
\end{minipage}
\begin{minipage}{.3\linewidth}
\centering  $\bm y_{t-2}$
\end{minipage}
\begin{minipage}{.3\linewidth}
\centering   Elements of $\bm{U}_t$
\end{minipage}\\
\begin{minipage}{.5\linewidth}
\includegraphics[scale=.45]{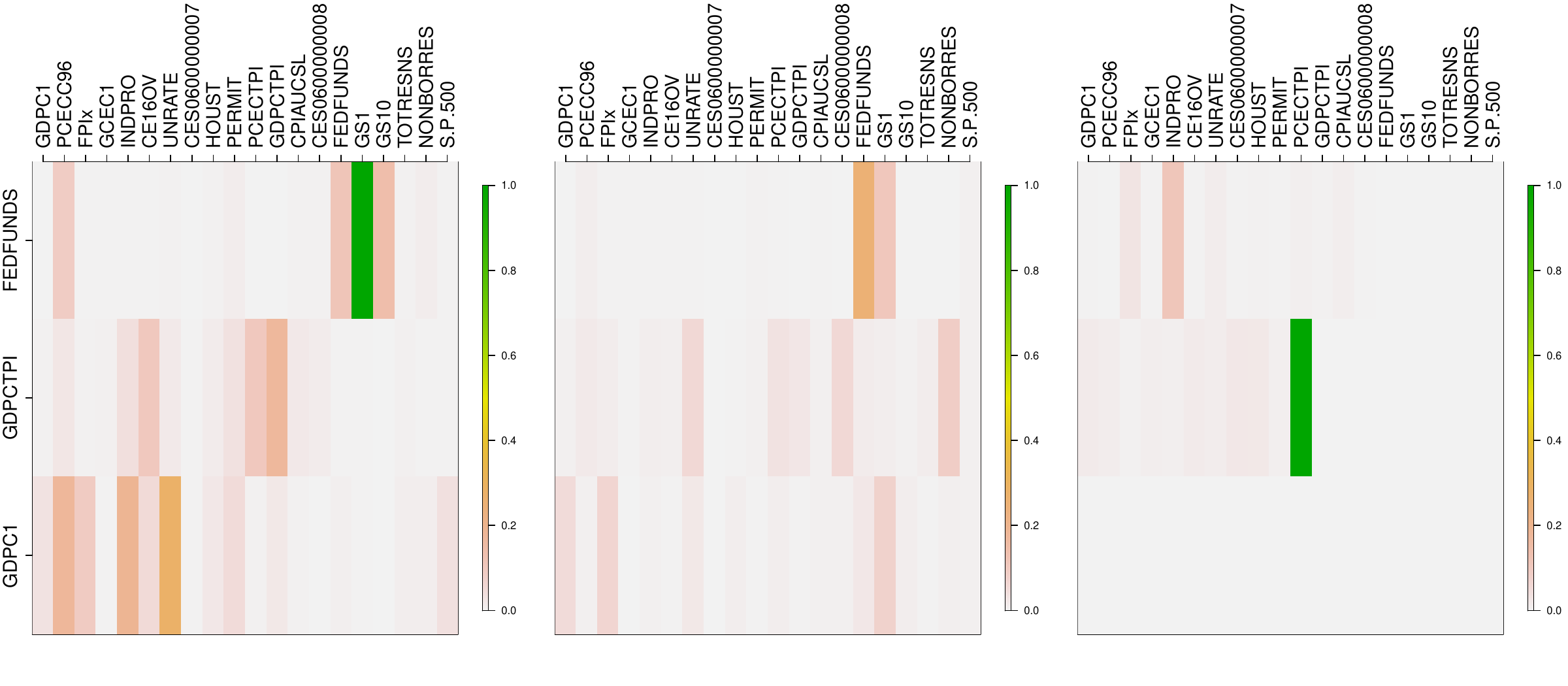}
\end{minipage}
\end{figure}

Before presenting the results of our forecasting exercise, we present Figure \ref{fig: heatmap} which sheds light on which variables our algorithm is choosing to predict the focus variables. This figure is produced using the large data set and the HS prior. Previously, we have discussed how doing sparsification for each MCMC draw shares similarities with Bayesian model averaging, allowing us to analyze PIPs. Figure \ref{fig: heatmap} is a heatmap of these PIPs at the end of the sample. Remember that, in the non-centered parameterization of the TVP-VAR-SV (see equation (\ref{eqeq: TVPVAR_noncen})), there are coefficients which appear on the initial states which are constant coefficients. The upper panel of the figure relates to these. The remaining coefficients determine whether there is time-variation relative to the constant coefficients. The lower half of the figure relates to these.

 Figure \ref{fig: heatmap} shows that our methods are inducing a high degree of sparsity in the TVP-VAR-SV in that most of the PIPs are near zero. However, a few of them are not. In terms of the VAR coefficients there is only one coefficient which is always selected (i.e. has a PIP of one). This is the first lag of the 1-year Treasury Bill rate in the equation for the Fed Funds rate. However, an appreciable number of other predictors have PIPs that are substantially above zero but much less than one.  In terms of the error covariance matrix, a similar pattern emerges. There is only one error covariance term which is non-zero in every MCMC draw.\footnote{Note that the other green areas refer to the diagonal elements of $\bm U_t$.} This is the covariance between the errors in the equations for two different inflation measures. However, there are several other error covariances with PIPs that are substantially above zero, even if they are below one. We stress that such a finding would not be possible if we were to use the SAVS algorithm directly on the posterior mean as opposed to using it on each MCMC draw. In the former case every PIP would be either zero or one with no values in between.

These patterns are consistent with those found in \cite{giannone2017economic} who conclude \textit{"there seems to be
a lot of uncertainty about whether certain predictors should be included in the model, which results into their selection only in a subset of the posterior draws. These findings reflect a substantial degree of collinearity among many predictors that carry similar information, hence complicating the task of structure discovery. In sum, model uncertainty is pervasive and the best prediction is obtained as a weighted average of several models." }These features seem to be exactly what our algorithm is uncovering in an automatic fashion. 

Finally, it is worth noting that there is evidence of time variation in several of the coefficients and our algorithm is automatically deciding which ones to allow to be time-varying. That is most of the PIPs which are appreciably above zero in the top half of the figure are also above zero in the bottom half. This pattern indicates a non-zero coefficient at time zero which is time varying. But our method also allows for a coefficient to be non-zero but constant. There are some cases which provide evidence of this. For instance, in the GDP growth equation the first lag of  S\&P500 stock returns has a PIP which is appreciably above zero in the top half of the figure, but is much closer to zero in the bottom half of the figure. This pattern indicates support for a constant coefficient on this predictor.

%\textcolor{blue}{Having written this, I realize I must have made a minor error in interpretation. I was assuming the PIPs are for the elements of the lower triangular $U_t$ when it is for the Cholesky decomposition. Please do the necessary re-writing if you want. However, I am wondering if it might be better to do $U_{t}$ since anyone looking at the figure will see the green jumping out in the far right part of the figure and think there is a lot of action going on in the error covariance matrix. But really all the green parts in the upper half are basically saying is that error variances are not zero (obvious) and the green parts in the bottom panel are saying the error variances are time varying (fairly obvious). The real story is that, with one exception, our approach is mostly shrinking error covarainces to zero and this would be better shown by a graph relating to  $U_t$? Or am I misinterpreting things?} 

The evidence in Figure \ref{fig: heatmap} suggests that shrinking then sparsifying is working in a sensible fashion. 
But the key test of our methodology is how well it forecasts. Table \ref{tab:LPS} presents the results of our forecasting exercise. A comparison of each set of sparsified forecasts to its non-sparsified counterpart shows the benefits of our shrink-then-sparsify strategy, particularly in large models. For $M=8$ and $M=20$, sparsification leads to substantial improvements in both RMSEs and LPLs in almost every case. These improvements are particularly noticeable for GDP forecasting for the one-step-ahead forecasts. In general, the benefits of sparsification are largest when using the DL or  Lasso priors. For $M=3$ the benefits of sparsification are less pronounced.   In terms of RMSEs, there seems to be no benefits of sparsification, although it does lead to slight improvements in the density forecasts even for this already fairly parsimonious case. This smaller accuracy premium from sparsification can be traced back to the fact that, in small models, {the detrimental influence of irrelevant but non-zero regression coefficients on predictive accuracy is small. In larger models, this effect eventually accumulates, leading to inflated posterior uncertainty and a decreased forecasting accuracy.
}
%increases in the predictive variance that arise from posterior uncertainty surrounding shrunk estimates are small relative to the variance contribution arising from the reduced-form shocks. In larger models, parameter uncertainty eventually accumulates and this seems to be detrimental for forecasting accuracy. 

In relation to the benchmark VAR-SV model, it is interesting to note that it is inferior to the TVP-VAR-SV models for the small and medium data sets. Clearly, addition of time-variation in the VAR coefficients helps improve forecasts in these cases. However, in the large data set, the evidence is mixed. In this case, the RMSEs produced by the TVP-VAR-SV are substantially better than those produced by the VAR-SV. However, the density forecasts are not. This could be due to the fact that there is typically a tradeoff between model dimension and parameter change. In small models, there is often a need for a high degree of parameter change to adequately fit patterns in the data and alleviate potential omitted variable biases. But in larger models, the information provided by the additional variables can fit these patterns, leaving less of a role for parameter change. Thus, in high dimensional cases the VAR-SV might be adequate and the extra flexibility provided by a TVP-VAR-SV may not be required. Of course, if the correct specification has a zero coefficient, the non-sparsified  approach would try and estimate the time-varying coefficient to be constant over time. But, as illustrated in Figure \ref{fig_ss}, estimation uncertainty   (although reduced) would still exist which could potentially hurt the forecasting performance of our approach. Sparsification as done in this paper clearly helps, but in the large data set there are still some cases where the VAR-SV is superior.  In such cases, a simple extension of our shrink-then-sparsify approach could help. In this paper, we have focused on sparsifying $\bm \alpha$ in equation (\ref{eq: tvp}). But any function of the parameters of a model could be sparsified in the same manner and, in particular, sparsifying the change in the states would be possible. This would lead to the constancy of a coefficient over certain periods in time while allowing for movements in other points in time when this kind of sparsification is applied. 

\begin{table}[t]
\caption{Relative root mean square errors (RMSEs) to a large BVAR-SV with a DL prior: 1997Q4 to 2017Q4}
{\tiny
\begin{center}
\begin{tabular}{llclllllclllll}
\toprule
\multicolumn{1}{l}{\bfseries }&\multicolumn{1}{c}{\bfseries }&\multicolumn{1}{c}{\bfseries }&\multicolumn{5}{c}{\bfseries Non-sparse}&\multicolumn{1}{c}{\bfseries }&\multicolumn{5}{c}{\bfseries Sparse}\tabularnewline
\cline{4-8} \cline{10-14}
\multicolumn{1}{l}{}&\multicolumn{1}{c}{}&\multicolumn{1}{c}{}&\multicolumn{1}{c}{DL}&\multicolumn{1}{c}{LASSO}&\multicolumn{1}{c}{NG}&\multicolumn{1}{c}{HS}&\multicolumn{1}{c}{MNIG}&\multicolumn{1}{c}{}&\multicolumn{1}{c}{DL}&\multicolumn{1}{c}{LASSO}&\multicolumn{1}{c}{NG}&\multicolumn{1}{c}{HS}&\multicolumn{1}{c}{MNIG}\tabularnewline
\midrule
{\scshape Small (m = 3)}&&&&&&&&&&&&&\tabularnewline
~~&\textbf{One-step-ahead}&&&&&&&&&&&&\tabularnewline
~~&GDPC1&&0.516&0.528&0.525&0.522&0.516&&0.525&0.529&0.531&0.532&0.525\tabularnewline
~~&&&(0.209)&(0.188)&(0.194)&(0.2)&(0.223)&&(0.228)&(0.231)&(0.227)&(0.222)&(0.229)\tabularnewline
~~&GDPCTPI&&0.861&0.866&0.866&0.864&0.871&&0.885&0.914&0.912&0.902&0.89**\tabularnewline
~~&&&(0.134)&(0.125)&(0.122)&(0.125)&(0.125)&&(0.114)&(0.075)&(0.076)&(0.085)&(0.105**)\tabularnewline
~~&FEDFUNDS&&0.839&0.856&0.843&0.834&0.834&&0.815&0.827&0.826&0.821&0.822\tabularnewline
~~&&&(0.495)&(0.466)&(0.509)&(0.547)&(0.554)&&(0.542**)&(0.564**)&(0.583**)&(0.604**)&(0.603**)\tabularnewline
~~&\textbf{Four-step-ahead}&&&&&&&&&&&&\tabularnewline
~~&GDPC1&&0.51&0.512&0.512&0.511&0.531&&0.545&0.542&0.545&0.547&0.547\tabularnewline
~~&&&(0.516)&(0.501)&(0.507)&(0.514)&(0.529)&&(0.542)&(0.553)&(0.545)&(0.547)&(0.537)\tabularnewline
~~&GDPCTPI&&0.983&0.983&0.986&0.986&0.983&&0.988&0.985&0.989&0.985&0.985*\tabularnewline
~~&&&(0.091)&(0.099)&(0.098)&(0.099)&(0.088)&&(0.081)&(0.072)&(0.071)&(0.073)&(0.079)\tabularnewline
~~&FEDFUNDS&&0.744&0.758&0.749&0.739&0.757&&0.761&0.738&0.737&0.745&0.763\tabularnewline
~~&&&(0.457)&(0.431)&(0.454)&(0.495)&(0.488)&&(0.466)&(0.468*)&(0.477*)&(0.505*)&(0.499**)\tabularnewline
\midrule
{\scshape Medium (m = 7)}&&&&&&&&&&&&&\tabularnewline
~~&\textbf{One-step-ahead}&&&&&&&&&&&&\tabularnewline
~~&GDPC1&&0.656&0.649&0.651&0.618&0.533&&0.492*&0.499&0.495*&0.489*&0.469*\tabularnewline
~~&&&(0.056)&(0.081)&(0.091)&(0.14)&(0.199)&&(0.237**)&(0.273**)&(0.273**)&(0.277*)&(0.273*)\tabularnewline
~~&GDPCTPI&&0.875&0.862&0.859&0.862&0.858&&0.88&0.911&0.909*&0.902*&0.877*\tabularnewline
~~&&&(0.106)&(0.126)&(0.134)&(0.137)&(0.147)&&(0.132)&(0.064)&(0.074)&(0.089)&(0.124)\tabularnewline
~~&FEDFUNDS&&0.849&0.868&0.855&0.819&0.83&&0.781&0.787&0.789&0.783&0.787\tabularnewline
~~&&&(0.273)&(0.327)&(0.408)&(0.552)&(0.464)&&(0.401)&(0.546**)&(0.581**)&(0.646**)&(0.593**)\tabularnewline
~~&\textbf{Four-step-ahead}&&&&&&&&&&&&\tabularnewline
~~&GDPC1&&0.62&0.624&0.621&0.594&0.533&&0.554&0.547&0.55&0.554&0.552\tabularnewline
~~&&&(0.22)&(0.272)&(0.297)&(0.377)&(0.438)&&(0.524**)&(0.545*)&(0.547*)&(0.533)&(0.513)\tabularnewline
~~&GDPCTPI&&0.992&0.98&0.98&0.979&0.988&&0.998&0.985&0.986&0.984&0.986\tabularnewline
~~&&&(-0.001)&(0.058)&(0.068)&(0.096)&(0.086)&&(0.07)&(0.025)&(0.033)&(0.054)&(0.067)\tabularnewline
~~&FEDFUNDS&&0.827&0.837&0.807&0.75&0.753&&0.771&0.739&0.729&0.736&0.749\tabularnewline
~~&&&(0.262)&(0.297)&(0.372)&(0.46)&(0.372)&&(0.417)&(0.443)&(0.46)&(0.448**)&(0.368**)\tabularnewline
\midrule
{\scshape Large (m = 20)}&&&&&&&&&&&&&\tabularnewline
~~&\textbf{One-step-ahead}&&&&&&&&&&&&\tabularnewline
~~&GDPC1&&1.001&0.835&0.834&0.844&0.693&&0.671**&0.546**&0.531**&0.53*&0.513*\tabularnewline
~~&&&(-0.355)&(-0.145)&(-0.119)&(-0.003)&(0.085)&&(-0.072**)&(0.212**)&(0.216**)&(0.214**)&(0.232**)\tabularnewline
~~&GDPCTPI&&1.119&1.051&1.089&1.018&0.929&&1.005**&0.941&1.029&0.952&0.912\tabularnewline
~~&&&(-0.054)&(0.048)&(0.032)&(0.062)&(0.084)&&(0.022**)&(0.041)&(0.024)&(0.03)&(0.054)\tabularnewline
~~&FEDFUNDS&&1.265&1.492&1.403&1.209&0.938&&1.025**&1.269**&1.23**&1.036**&0.858**\tabularnewline
~~&&&(-0.555)&(-0.443)&(-0.326)&(0.2)&(0.154)&&(-0.32**)&(-0.119**)&(-0.005**)&(0.516**)&(0.552**)\tabularnewline
~~&\textbf{Four-step-ahead}&&&&&&&&&&&&\tabularnewline
~~&GDPC1&&0.99&0.554&0.515&0.497&0.538&&0.655*&0.539&0.535&0.542&0.57\tabularnewline
~~&&&(-1.189)&(-0.316)&(-0.244)&(0.188)&(0.363)&&(-0.338**)&(0.536**)&(0.513**)&(0.56**)&(0.543**)\tabularnewline
~~&GDPCTPI&&1.067&1.059&1.074&1.015&0.989&&1.016**&1.006&1.041&0.992*&0.986\tabularnewline
~~&&&(-0.815)&(-0.184)&(-0.149)&(0.01)&(0.03)&&(-0.287**)&(-0.022)&(-0.03)&(0.012)&(0.012)\tabularnewline
~~&FEDFUNDS&&0.913&0.871&0.826&0.784&0.84&&0.804&0.771&0.759*&0.786&0.825\tabularnewline
~~&&&(-1.456)&(-0.714)&(-0.602)&(0.085)&(0.094)&&(-0.729**)&(0.02**)&(0.056**)&(0.274)&(0.223*)\tabularnewline
\bottomrule
  \multicolumn{14}{p{18.3cm}}{\vspace{.025cm}\footnotesize{\textbf{Notes:} Numbers in parentheses refer to the average log predictive likelihoods (LPLs) vis-\'{a}-vis the BVAR-SV with a DL prior. DL refers to a TVP-VAR-SV with a  Dirichlet-Laplace prior, Lasso to the Bayesian Lasso, NG to the Normal-Gamma prior, HS to the Horseshoe, and NMIG to the Normal-mixture of Inverse Gamma prior. Asterisks indicate statistical significance between a sparsified model and its non-sparsified model at the $5$ ($^{**}$) and $10$ ($^{*}$) percent significance level.}}
\end{tabular}

\label{tab:LPS}
\end{center}}
\end{table}
{The results in Table \ref{tab:LPS} highlight that, when the full hold-out period is considered, sparsification often improves predictive accuracy relative to a non-sparsified model specification. In selected cases, however, the SAVS step also seems to hurt forecasting accuracy. This raises the question whether  a practitioner should  always use the sparsification step.  The findings in the table indicate that  relative accuracy gains obtained from using SAVS increase with model size. In large models (with $M=20$),  improvements in predictive accuracy  for point and density forecasts are often substantial (see, e.g., the improvements for GDP growth and short-term interest rates) while in small models, differences are often negligible and sometimes favor non-sparsified models.   This suggests that in large models, using SAVS appears to improve forecasts. However, we would like to stress that applying SAVS  yields predictions that are always competitive relative to non-sparsified competitors, even in small-dimensional models. This is corroborated by  \cite{diebold1995paring} tests which suggest that in most cases where the SAVS step lowers predictive accuracy, these decreases are statistically insignificant whereas in the case that sparsification improves forecast accuracy, the differences are often significant. Thus, we can recommend applying the sparsification step in all large-dimensional cases (i.e. for $M\ge 7$) since the computational burden is not increased significantly while the predictive performance is adversely affected in only a few situations in a significant manner. }

\begin{figure}[h!]
\caption{Performance differences between a sparsified and non-sparsified TVP-VAR-SV with a HS prior}\label{fig: BF_spars_shrink}
\begin{minipage}{\linewidth}
\vspace{.3cm}
\centering (a) Evolution of log predictive Bayes factor (sparse versus non-sparse)
\end{minipage}
\begin{minipage}{.33\linewidth}\vspace{.3cm}
\centering \textit{Small}
\end{minipage}
\begin{minipage}{.33\linewidth}\vspace{.3cm}
\centering \textit{Medium}
\end{minipage}
\begin{minipage}{.33\linewidth}\vspace{.3cm}
\centering \textit{Large}
\end{minipage}\\
\begin{minipage}{.33\linewidth}
\includegraphics[scale=.35]{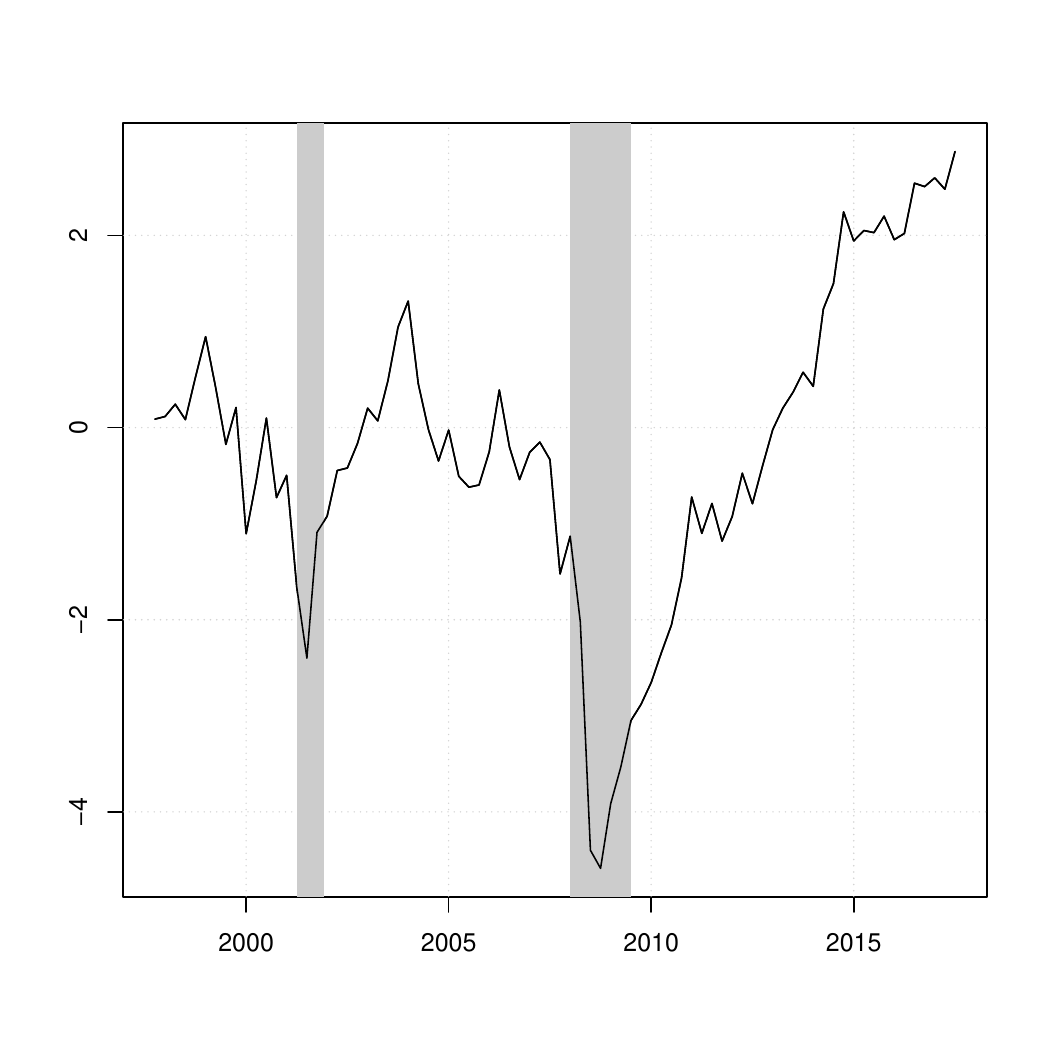}
\end{minipage}
\begin{minipage}{.33\linewidth}
\includegraphics[scale=.35]{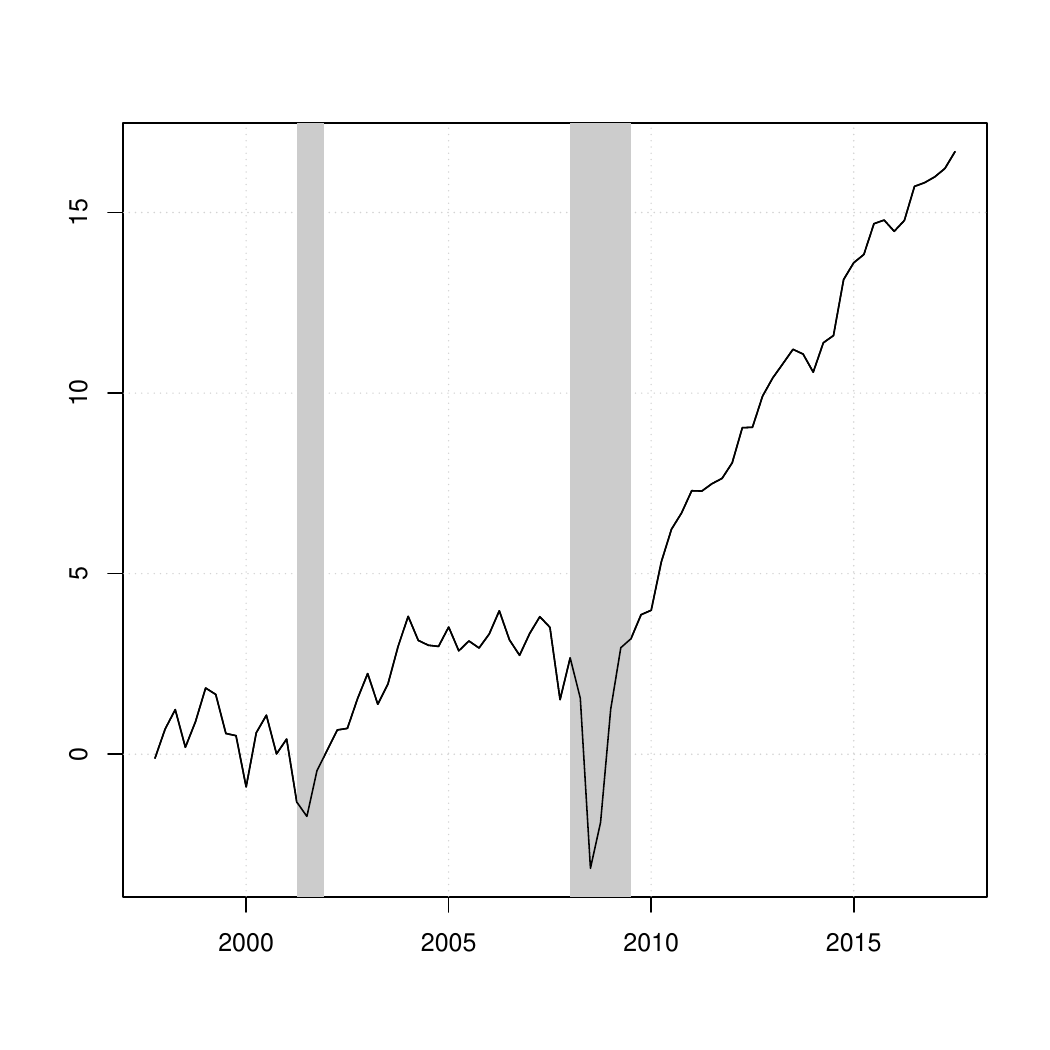}
\end{minipage}
\begin{minipage}{.33\linewidth}
\includegraphics[scale=.35]{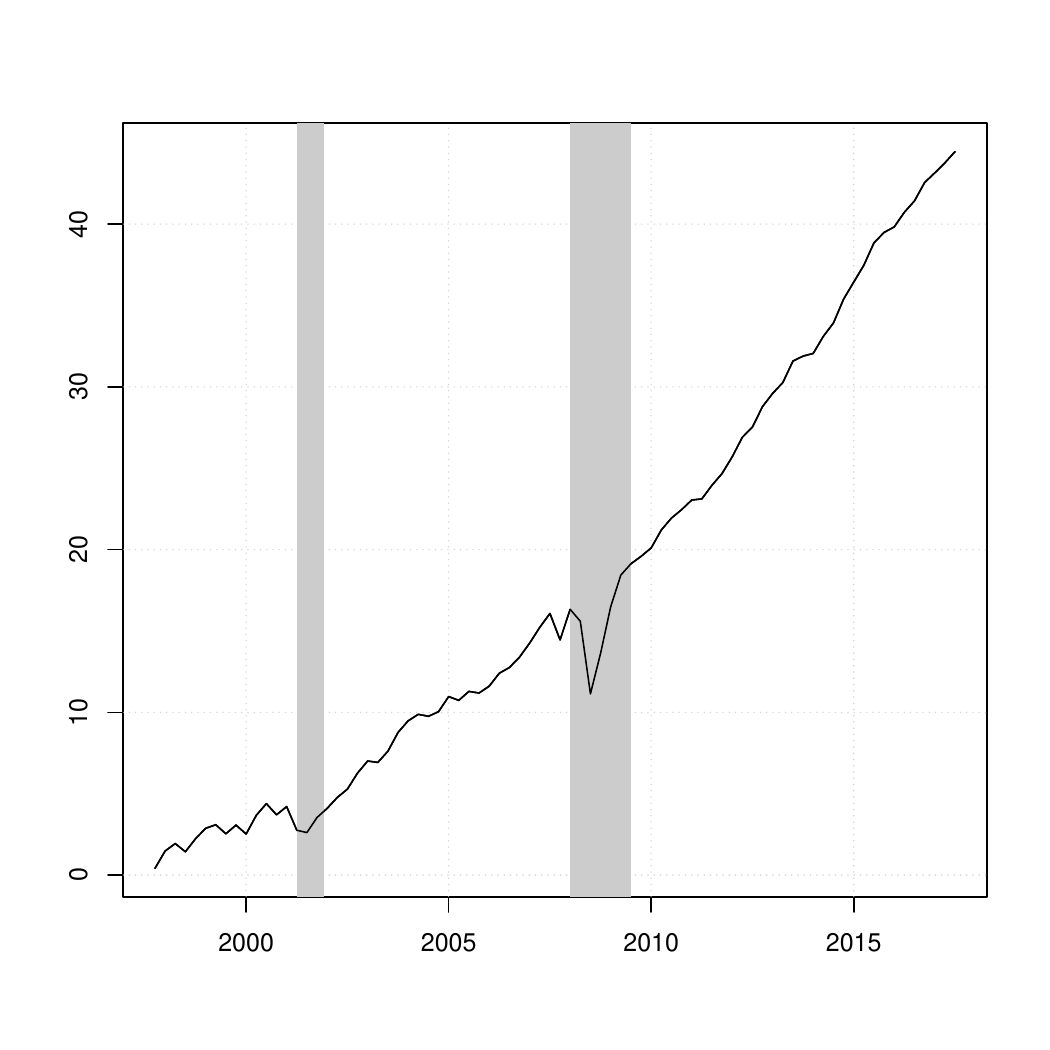}
\end{minipage}\\
\begin{minipage}{\linewidth}
\centering (b) Evolution of cumulative squared forecast errors
\end{minipage}
\begin{minipage}{.33\linewidth}\vspace{.3cm}
\centering \textit{Small}
\end{minipage}
\begin{minipage}{.33\linewidth}\vspace{.3cm}
\centering \textit{Medium}
\end{minipage}
\begin{minipage}{.33\linewidth}\vspace{.3cm}
\centering \textit{Large}
\end{minipage}\\
\begin{minipage}{.33\linewidth}
\includegraphics[scale=.35]{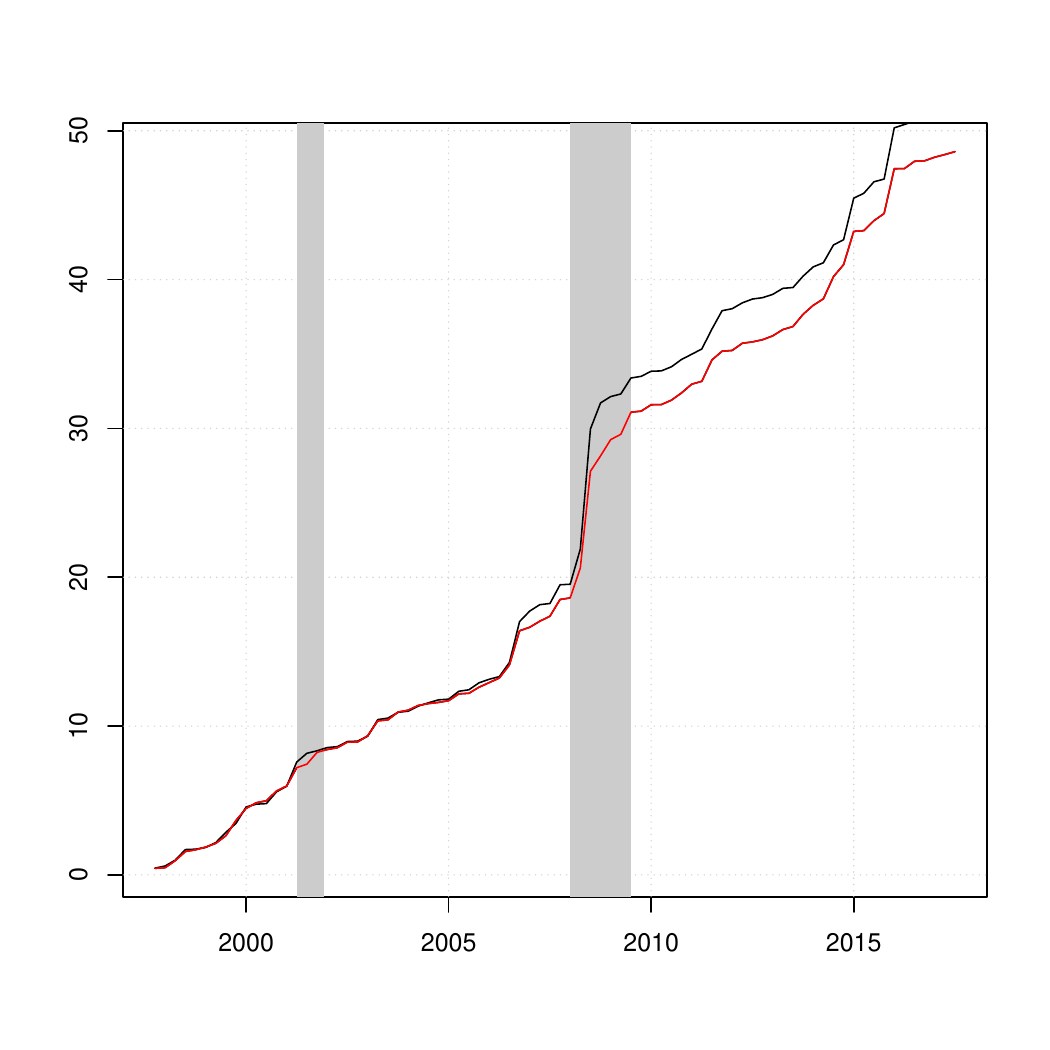}
\end{minipage}
\begin{minipage}{.33\linewidth}
\includegraphics[scale=.35]{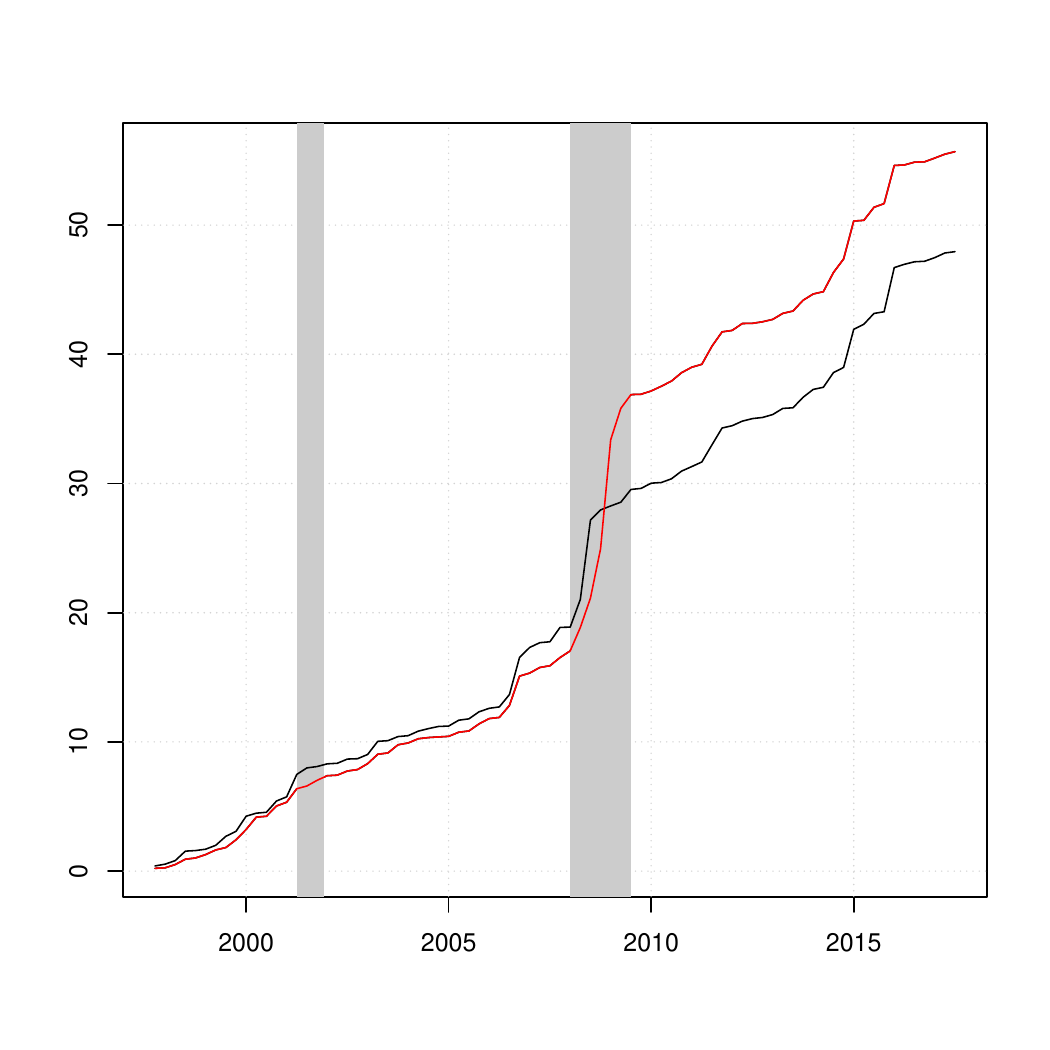}
\end{minipage}
\begin{minipage}{.33\linewidth}
\includegraphics[scale=.35]{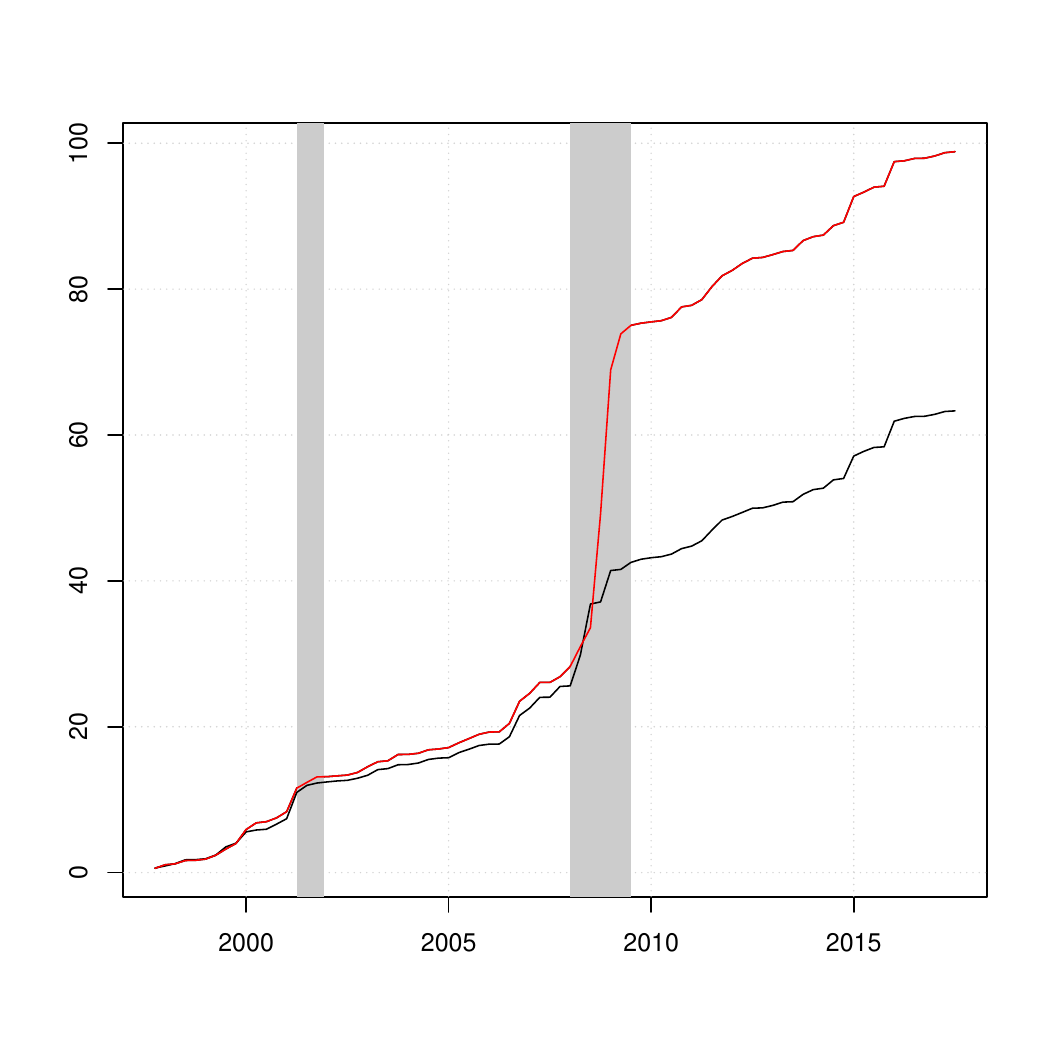}
\end{minipage}
\begin{minipage}{\linewidth}
\textbf{Notes}: 
The log-predictive Bayes factor between the sparsified  and non-sparsified model is obtained by considering the joint one-step-ahead predictive density for the three focus variables and the squared forecast errors are averages across the one-step-ahead forecast errors for the focus variables.  The black line in panel (b) refers to the sparsified squared forecast error while the red line denotes the non-sparsified model.  The gray shaded areas refer to NBER reference recessions in the US. 
\end{minipage}
\end{figure}

The discussion in the previous paragraph provides a simple recommendation that is based on using the full hold-out period.
In the next step, we ask whether  accuracy differences could also be specific to certain periods in time. To this end, Figure \ref{fig: BF_spars_shrink}(a)  shows the evolution of the log predictive Bayes factor between the sparsified and non-sparsified large-scale TVP-VAR-SV with the HS prior over the hold-out period.\footnote{Comparable figures for other shrinkage priors reveal similar patterns. Thus. for the sake of brevity, we discuss the results for the HS prior exclusively.} This Bayes factor is obtained by evaluating the one-step-ahead predictive density for the three focus variables jointly after integrating out the remaining variables. To investigate whether the gains in density forecasting performance stem from capturing higher order moments in the predictive distribution or from a more precise point forecast, Figure \ref{fig: BF_spars_shrink}(b) shows cumulative squared one-step-ahead forecast errors averaged across the focus variables over time.

Figure \ref{fig: BF_spars_shrink}(a)  indicates that accuracy premia from sparsification tend to vary significantly over the business cycle. During expansionary stages, sparsification yields modest (in the case of the medium-sized model) to sustained (in the case of the large model) improvements in density forecasting performance relative to the non-sparsified competitor. For the small-scale TVP-VAR-SV, accuracy gains are more muted during expansionary periods. During recessions, in contrast, sparse models tend to be outperformed by their non-sparsified counterparts. Our conjecture is that this stems from the fact that during turbulent times, the sparsified predictive distributions feature a smaller variance, making it harder to capture outlying observations and thus translating in lower log predictive likelihoods.

Our conjecture is confirmed when focusing on point forecasts. In terms of point predictions, we observe that forecast errors are almost identical in the period up to the global financial crisis. During the recession in 2008/2009, forecast errors increase markedly but slightly less so for the sparsified model and for the medium and large dataset. This suggests that the drop in the log predictive Bayes factor is mainly driven by higher order moments, implying that while the accuracy of the point prediction increases, adverse movements in the corresponding predictive variance offset this gain.

\section{Conclusions}\label{concl}

Global-local shrinkage priors have enjoyed great popularity in over-parameterized regressions and VARs involving large numbers of variables. And, increasingly, they have been used with TVP versions of these models which are potentially even more over-parameterized. Use of such priors can potentially reduce estimation error and improve forecasts. However, estimation error is not completely eliminated and it is possible that further improvements in forecasting performance can be achieved by adding an additional sparsification step to shrunk estimates to further reduce the lower bound on accuracy associated with shrinkage. In this paper, we have developed methods for doing so. Compared to a recent paper, \cite{woody2019}, who develop methods for quantifying posterior uncertainty in sparsified models based on using a single sparse model, our approach controls for model uncertainty by sparsifying each MCMC draw. In an artificial data exercise, we have shown that our shrink-then-sparsify approach to TVP regression leads to more accurate estimates for a variety of DGPs. Particularly large gains are found in sparse DGPs. In a macroeconomic forecasting exercise, adding sparsification to shrinkage also leads to substantial improvements in forecast performance if interest centers on using large models.

\newpage
%\addcontentsline{toc}{section}{References}
%\bibliographystyle{plain}
%\bibliography{sparse}
\bibliographystyle{fischer}
\bibliography{sparse}

\clearpage
\begin{appendices}
\section{Global-local Shrinkage Priors}\label{sec: GL_priors}
The first four sub-sections of this appendix provide relevant details on the prior setup, briefly discussing key features of the used priors, hyperparameter choices used, and relevant information necessary to perform posterior inference. 
\subsection{The Dirichlet-Laplace Prior}
The DL prior, originally proposed in \cite{bhattacharya2015dirichlet}, assumes that each element in $\bm \alpha$, $\alpha_j~(j=1,\dots, 2K)$, follows a Gaussian distribution,
\begin{equation*}
\alpha_j|\omega_j, \xi_j,\lambda \sim \mathcal{N}(0, \omega_j \xi^2_j  \zeta^2),
\end{equation*}
with
\begin{align*}
\omega_j \sim \mathcal{E}(1/2), \quad
\xi_j \sim \mathcal{D}(a, \dots,a), \quad
\zeta \sim \mathcal{G}\left( 2Ka, 1/2\right)
\end{align*}
where  $\omega_j$ is a variable-specific scaling parameter that features an exponentially distributed prior, with $\mathcal{E}$ denoting the exponential distribution, $\xi_j$ denotes yet another local shrinkage parameter with $\bm \xi = (\xi_1, \dots, \xi_{2K})'$ being bounded to the $(2K-1)$-dimensional simplex (i.e. $\xi_j \ge 0$ and $\sum_{j}\xi_j=1$). We use a Dirichlet distributed prior with intensity parameter $a$ on $\xi_j$.  Finally,  $\zeta$ is a global shrinkage term that follows a Gamma distribution. Notice that the relationship between this prior hierarchy and the general form  provided in equation (\ref{eq: GL_general}) can be seen by defining $\phi_j = \omega_j \xi_j^2$ and $\lambda=\zeta^2$.

\cite{bhattacharya2015dirichlet} show within the stylized normal means problem that the optimal value of $a$ is specified to be $(2K)^{-(1+\varphi)}$ with $\varphi$ being a positive number close to zero. Since this hyperparameter plays a crucial role in determining the shrinkage behavior of the DL prior, we estimate it using a prior which is a  uniform distribution that is bounded between $(2K)^{-1}$ and $1/2$.

Posterior simulation can be carried out using a slightly modified variant of the MCMC algorithm proposed in \cite{bhattacharya2015dirichlet}. The full conditional posterior distribution of $\omega_j$ follows an  inverse Gaussian distribution:
\begin{equation*}
\omega_j |\alpha_j, \xi_j, \zeta \sim \mathcal{iG}\left(\zeta \frac{\xi_j}{|\alpha_j|}, 1\right).
\end{equation*}
The global shrinkage parameter $\zeta$ follows a generalized inverted Gaussian (GIG) distribution,
\begin{equation*}
\zeta | \bm \alpha, \bm \xi \sim \mathcal{GIG}\left(2K (a-1), 1, 2 \sum_{j=1}^{2K}\frac{|\alpha_j|}{\xi_j}\right).
\end{equation*}
Moreover, we draw the second set of local scaling parameters $\xi_j$ by introducing auxiliary variables $T_j$ that follow a GIG distribution:
\begin{equation*}
T_j|a, \alpha_j \sim \mathcal{GIG}(a-1, 1, 2 |\alpha_j|).
\end{equation*}
We then set $\xi_j = T_j / \sum_{i=1}^{2K} T_i$ to obtain a valid draw from the full conditional posterior of $\xi_j$. 

To simulate from the conditional posterior of $a$, we employ a Metropolis Hastings algorithm with a  Gaussian proposal distribution truncated between $(2K)^{-1}$ and $1/2$. The variance of the proposal distribution is tuned during the first $20$ percent of the burn-in stage of the MCMC sampler  such that the acceptance rate is between $20$ and $40$ percent.  

\subsection{The Normal-Gamma Prior and the Lasso}
As compared to the DL prior, the NG prior proposed in \cite{griffin2010inference} consists of a single group of idiosyncratic scaling factors $\phi_j$ and a global shrinkage parameter $\lambda=1/\tilde{\lambda}$. We assume that each $\alpha_j$ follows a zero mean Gaussian distribution a priori:
\begin{equation*}
\alpha_j | \phi_j, \tilde{\lambda} \sim \mathcal{N}(0, \phi_j), \quad \phi_j|\tilde{\lambda} \sim \mathcal{G}(\vartheta, \vartheta \tilde{\lambda}/2), \quad \tilde{\lambda} \sim \mathcal{G}(d_{\tilde{\lambda}},e_{\tilde{\lambda}}).
\end{equation*}
Here, we let $\vartheta$ denote a hyperparameter that controls the tail behavior of the prior, with smaller values of $\vartheta$ leading to heavier tails and  increasing mass is placed on zero while larger value do the opposite.  $d_{\tilde{\lambda}}$ and $e_{\tilde{\lambda}}$ are hyperparameters that control the overall degree of shrinkage, with values close to zero implying heavy shrinkage towards zero.

One key feature of the NG prior is that it nests the Bayesian Lasso of \cite{park2008bayesian} by setting $\vartheta=1$. Since $\vartheta$ plays a crucial role, we follow \cite{griffin2010inference} and introduce an Exponential prior on $\vartheta$:
\begin{equation*}
\vartheta \sim \mathcal{Exp}(\underline{\vartheta}).
\end{equation*}
$\underline{\vartheta}$ is set equal to $1$, pushing the prior towards the Bayesian Lasso. Moreover, we set $d_{\tilde{\lambda}}=e_{\tilde{\lambda}}=10^{-4}$, implying a disperse prior on $\tilde{\lambda}$ and thus being consistent with heavy shrinkage (by allowing large values of $\tilde{\lambda}$).

The hierarchical structure of the prior yields closed-form full conditionals for $\phi_j$ and $\tilde{\lambda}$. The local scaling parameters $\phi_j$ follow a GIG distribution:
\begin{equation*}
\phi_j | \tilde{\lambda}, \alpha_j \sim \mathcal{GIG}\left(\vartheta-\frac{1}{2}, \vartheta \tilde{\lambda}, \alpha_j^2\right).
\end{equation*}
For the global shrinkage parameter, we obtain a Gamma-distributed full conditional posterior distribution:
\begin{equation*}
\tilde{\lambda}|\phi_1, \dots, \phi_{2K}, \vartheta \sim \mathcal{G}\left(d_{\tilde{\lambda}}+\vartheta 2K, e_{\tilde{\lambda}}+ \frac{\vartheta}{2} \sum_{i=1}^{2K} \phi_i\right).
\end{equation*}
Finally, we obtain draws from the conditional posterior of $\vartheta$ by setting up a random walk MH algorithm in terms of $\log \vartheta$ \citep[see][]{griffin2010inference}. 

\subsection{The Horseshoe Prior}
For the HS prior of \cite{carvalho2010horseshoe}, we consider the representation based on auxiliary variables proposed in \cite{makalic2016simple}. The corresponding prior hierarchy is given by
\begin{align*}
\alpha_j | \lambda, \phi_j \sim \mathcal{N}(0, \phi_j \lambda), \quad \phi_j \sim \mathcal{G}^{-1}(1/2, 1/\nu_j), \quad \lambda \sim \mathcal{G}^{-1}(1/2, 1/\varphi),
\end{align*}
whereby $\nu_j$ and $\varphi$ denote auxiliary variables and $\mathcal{G}^{-1}$ denotes the inverse Gamma distribution. The auxiliary variables also follow inverse Gamma distributions,
\begin{equation*}
\nu_1, \dots, \nu_{2K}, \varphi 	\sim \mathcal{G}^{-1}(1/2, 1).
\end{equation*}
This representation of the HS prior allows for straightforward updating of the local and global scaling parameters and involves sampling from inverted Gamma distributions exclusively. The corresponding full conditional posterior distributions are
\begin{align*}
\phi_j| \alpha_j, \lambda, \nu_j &\sim \mathcal{G}^{-1}\left(1, \frac{1}{\nu_j}+\frac{\alpha_j^2}{2 \lambda}\right)\\
\lambda|\alpha_j, \phi_j, \varphi &\sim \mathcal{G}^{-1}\left(\frac{2K+1}{2}, \frac{1}{\varphi}+\frac{1}{2}\sum_{i=1}^{2K} \frac{\alpha^2_i}{\phi_j}\right).
\end{align*}
The conditional posteriors of the auxiliary variables are given by
\begin{align*}
\nu_j|\phi_j \sim \mathcal{G}^{-1}\left(1, 1+ \frac{1}{\phi_j^2}\right), \\
\varphi|\lambda \sim \mathcal{G}^{-1}\left(1, 1+\frac{1}{\lambda}\right).
\end{align*}

\subsection{The Normal-mixture of Inverse Gamma Prior}
The NMIG prior of \cite{ishwaran2005spike} extends the original SSVS prior proposed in \cite{george1993variable, george1997approaches} along several dimensions. To set the stage, we use
a  mixture of Gaussians prior distribution on $\alpha_j$:
\begin{equation*}
\alpha_j | \delta_j, \tau^2_j \sim \mathcal{N}(0, \tau^2_j)  \delta_j + \mathcal{N}(0, c \tau^2_j) (1-\delta_j),
\end{equation*}
where $\delta_j$ denotes a Bernoulli random variable with prior probability $Prob(\delta_j =1)= \underline{p}$ while $c$ is a constant close to zero and $\tau^2_j$ is a coefficient-specific scaling factor. Following \cite{ishwaran2005spike}, we specify an inverse Gamma prior on $\tau^2_j$ and a Beta distributed prior on $\underline{p} $:
\begin{align*}
\tau^2_j  &\sim \mathcal{G}^{-1}(d_\tau, e_\tau),\\
\underline{p} &\sim \mathcal{B}(d_p, e_p),
\end{align*}
with $d_\tau, e_\tau, d_p$ and $e_p$ denoting hyperparameters. Notice that this specification implies conditional prior independence between the indicators $\delta_j$. However, the common prior inclusion probability $\underline{p}$ serves as a common factor, implying that marginally, the indicators are dependent. 

\cite{ishwaran2005spike} notice that after integrating out $\tau_j^2$ and $\underline{p}$, the two components in the prior follow $t$-distributions. The hyperparameter $d_\tau$ controls the degrees of freedom of the marginal prior while the variances are given by $c e_\tau/d_\tau$ (for the spike component) and $e_\tau/d_\tau$ (for the slab component). In the empirical applications, we set $e_p = d_p = 1$, implying a Uniform prior on $\underline{p}$ and $c= 2.5/10^5$. Moreover, we set $d_\tau = 5$, leading to $10$ degrees of freedom and $e_\tau=4$. This is the benchmark prior setup as specified in \cite{malsiner2011comparing}.

For this prior specification, all conditional posterior distributions are available in closed form. The full-conditional posterior of $\delta_j$ follows a Bernoulli distribution with posterior probability $\overline{p}_j$ given by:
\begin{equation*}
\overline{p}_j = Prob(\delta_j = 1| \alpha_j, \tau_j^2, \underline{p})= \frac{\frac{1}{\tau^2_j} \text{exp}\left(-\frac{1}{2} \frac{\alpha^2_j}{\tau^2_j}\right)}{\frac{1}{\tau^2_j} \text{exp}\left(-\frac{1}{2} \frac{\alpha^2_j}{\tau^2_j}\right) \underline{p} + \frac{1}{c \tau^2_j} \text{exp}\left(-\frac{1}{2} \frac{\alpha^2_j}{c \tau^2_j}\right) (1-\underline{p})}.
\end{equation*}
The scaling factors $\tau_j^2$ follow an inverted Gamma distribution 
\begin{equation*}
\tau_j^2|\alpha_j^2, \delta_j \sim \mathcal{G}^{-1}\left( d_\tau + \frac{1}{2}, e_\tau + \frac{\alpha^2_j}{\delta_j + (1-\delta_j) c}\right).
\end{equation*}
Finally, the posterior distribution of $\underline{p}$ is a Beta distribution:
\begin{equation*}
\underline{p}| \delta_1, \dots, \delta_{2K} \sim \mathcal{B}\left(d_p + \sum_{j=1}^{2K} \delta_j, e_p + 2K - \sum_{j=1}^{2K} \delta_j\right).
\end{equation*}

\section{Full Conditional Posterior Simulation} \label{sec: MCMC}
For the dynamic regression models used in the main body of the text we use a relatively standard MCMC algorithm. Since we estimate the TVP-VAR-SV on an equation-by-equation basis, we describe the MCMC algorithm for the TVP regression model only. However, it is worth noting that all priors described in the previous subsection are specified to be equation-specific. This implies that instead of having a single global shrinkage parameter $\lambda$, each equation features its own global (equation-specific) shrinkage parameter. Moreover, one additional difference is that the dynamic regression model in Section \ref{ss_tvp} features homoscedastic errors. In the TVP-VAR case, we allow for stochastic volatility, implying that the MCMC algorithm differs slightly.

Our posterior simulator cycles between the following steps:
\begin{enumerate}
\item Simulate the full-history of $\tilde{\bm \beta}_t$, conditional on the remaining parameters, using the forward-filtering backward-sampling algorithm proposed in \cite{carterkohn} and \cite{fs1994} while exploiting the non-centered parameterization.
\item Sample the error variances from an inverted Gamma full conditional posterior distribution:
\begin{equation*}
\sigma_\varepsilon^2|\bullet \sim \mathcal{G}^{-1}\left(d_\sigma+T/2, e_\sigma+\frac{1}{2}\sum_{t=1}^T (y_t - \bm \alpha' \bm Z_t)^2\right).
\end{equation*}
where the $\bullet$ indicates conditioning on all parameters and the data.

\item Conditional on $\{\tilde{\bm \beta}_t\}_{t=1}^T$ and $\sigma^2_\varepsilon$, the conditional posterior of $\bm \alpha$ takes a multivariate Gaussian form:
\begin{equation*}
\bm \alpha|\bm \bullet \sim \mathcal{N}(\overline{\bm \alpha},  \overline{\bm \Omega}),
\end{equation*}
with
\begin{align*}
\overline{\bm \Omega}  &= (\sigma^{-2}_\varepsilon \bm Z' \bm Z + \underline{\bm \Omega}^{-1})^{-1},\\
\overline{\bm \alpha} &= \overline{\bm \Omega} \left(\sigma^{-2}_\varepsilon  \bm Z' \bm y\right)
\end{align*}
where $\bm Z$ is a $T \times 2K$ matrix with the $t^{{th}}$row equal to $\bm Z'_t$. Likewise, $\bm y= (y_1, \dots, y_T)'$ is a $T$-dimensional vector. $\bm \Omega$ denotes a diagonal prior variance-covariance matrix with typical element depending on the specific shrinkage prior chosen.

\item Depending on the global-local shrinkage prior adopted, construct the matrix $\bm \Omega$  based on the conditional posterior distributions outlined in Appendix \ref{sec: GL_priors}.
\end{enumerate}
In case we use a stochastic volatility specification for the error variances, we use the algorithm proposed in \cite{kastner2014ancillarity} and implemented in the R package stochvol \citep{kastner2016dealing}.  For the VAR case, the main steps of this algorithm remain identical except that the different steps of the algorithm can be interpreted as being specific to a given equation of the model. In all applications, we repeat this algorithm $30,000$ times and discard the first $15,000$ draws as burn-in.

\newpage
\section{Data Appendix}
\begin{table}[h!]
{\tiny
\begin{center}
\begin{tabular}{lllrlll}
\toprule
\multicolumn{1}{l}{\ }&\multicolumn{1}{c}{\ FRED Mnemonic}&\multicolumn{1}{c}{\ Description}&\multicolumn{1}{c}{\ Transformation Codes}&\multicolumn{1}{c}{\ SMALL}&\multicolumn{1}{c}{\ MEDIUM}&\multicolumn{1}{c}{\ LARGE}\tabularnewline
\midrule
{\scshape }&&&&&&\tabularnewline
~~&GDPC1&Real Gross Domestic Product&$5$&x &x &x\tabularnewline
~~&PCECC96&Real Personal Consumption Expenditures&$5$&&x&x\tabularnewline
~~&FPIx&Real private fixed investment &$5$&&x&x\tabularnewline
~~&GCEC1&Real Government Consumption Expenditures and Gross Investment&$5$&&&x\tabularnewline
~~&INDPRO&IP:Total index Industrial Production Index (Index 2012=100)&$5$& & &x\tabularnewline
~~&CE16OV&Civilian Employment (Thousands of Persons)&$5$&&x&x\tabularnewline
~~&UNRATE&Civilian Unemployment Rate (Percent)&$2$& & &x\tabularnewline
~~&CES0600000007&Average Weekly Hours of Production and Nonsupervisory Employees:  Goods-Producing&$2$&&x&x\tabularnewline
~~&HOUST&Housing Starts: Total: New Privately Owned Housing Units Started&$5$&&&x\tabularnewline
~~&PERMIT&New Private Housing Units Authorized by Building Permits&$5$&&&x\tabularnewline
~~&PCECTPI&Personal Consumption Expenditures: Chain-type Price Index &$6$&&&x\tabularnewline
~~&GDPCTPI&Gross Domestic Product: Chain-type Price Index&$6$&x&x&x\tabularnewline
~~&CPIAUCSL&Consumer Price Index for All Urban Consumers:  All Items&$6$&&&x\tabularnewline
~~&CES0600000008&Average Hourly Earnings of Production and Nonsupervisory Employees:&$6$&&x&x\tabularnewline

~~&FEDFUNDS&Effective Federal Funds Rate (Percent)&$2$&x&x&x\tabularnewline

~~&GS1&1-Year Treasury Constant Maturity Rate (Percent)&$2$&&&x\tabularnewline
~~&GS10&10-Year Treasury Constant Maturity Rate (Percent)&$2$&&&x\tabularnewline
~~&TOTRESNS&Total Reserves of Depository Institutions &$6$&&&x\tabularnewline
~~&NONBORRES&Reserves Of Depository Institutions, Nonborrowed&$7$&&&x\tabularnewline
~~&S.P.500&S\&P’s Common Stock Price Index:  Composite&$5$&&&x\tabularnewline
\bottomrule
\multicolumn{7}{p{18.3cm}}{\footnotesize{\textbf{Notes:} Transformation codes for a series $y$: $(1)$ no transformation, $(2)$  first differences $\delta y_t$, $(3)$ double differences $\Delta^2 y_t$  $(4)$  logarithmic transform $\log y_t$, $(5)$ difference of the logarithm $\Delta \log y_t$, $(6)$ double difference of the logarithm $\Delta^2 \log y_t$, and $(7)$ $\Delta (y_t / y_{t-1} -1)$.}}
\end{tabular}
\caption{Data description\label{tab:data-descr1}}\end{center}}
\end{table}
\end{appendices}

\end{document}